# A Unified Theory on Construction and Evolution of the Genetic Code

**Liaofu Luo***

**Laboratory of Theoretical Biophysics, Faculty of Physical Science and Technology,**

**Inner Mongolia University, Hohhot 010021, China**

*Email address: lolfcm@mail.imu.edu.cn

## Abstract

A quantitative theory on the construction and the evolution of the genetic code is proposed. Through introducing the concept of mutational deterioration (MD) and developing a theoretical formalism on MD minimization we have proved: 1，the redundancy distribution of codons in the genetic code obeys MD minimization principle; 2, the hydrophilic-hydrophobic distribution of amino acids on the code table is global MD (GMD) minimal; 3, the standard genetic code can be deduced from the adaptive minimization of GMD; 4, the variants of the standard genetic code can be explained quantitatively by use of GMD formalism and the general trend of the evolution is GMD non-increasing which reflects the selection on the code. We have demonstrated that the redundancy distribution of codons and the hydrophobic-hydrophilic (H-P) distribution of amino acids are robust in the code relative to the mutational parameter, and indicated that the GMD can be looked as a non-fitness function on the adaptive landscape. Finally, an important aspect on the symmetry of the code construction, the Yin-Yang duality is investigated. The Yin-Yang duality among codons affords a sound basis for understanding the H-P structure in the genetic code.

The approximate universality of the canonical genetic code and the discoveries of various deviant codes in a wide range of organisms strongly reveal that the genetic code is still evolving. Several mechanisms on code evolution were proposed, for example, the codon capture and the ambiguous decoding by tRNA [Knight et al, 2001; Santos et al, 2004]. However, a unified theory still lacks for a full explanation of the genetic code evolution both in its high universality and various deviations. Evidently, the point is closely related to the construction of the code. The construction of the genetic code obeys some general rules that afford a basis for understanding the universality and changeability of the code. On the other hand, the error minimization property of the genetic code was analyzed by several authors [Di Giuilo et al, 1994; Freeland & Hurst, 1998]. But it is still unclear why the canonical genetic code takes the standard form with error non-minimized and what are the evolutionary constraints for deducing the standard code. In the article we emphasize the unified understanding of the code construction and code evolution. We shall indicate that the unification between code construction and code evolution can be achieved through introducing the concept of mutational deterioration (MD) and developing a theoretical formalism for MD minimization. The materials are organized in the article as follows. In the first section we will review the mutational deterioration theory on the redundancy distribution in the genetic code. Then the adaptive minimization of global mutational deterioration and the accuracy of the genetic code will be discussed in the second section. Next, in the third section, we will study the evolvability of the genetic code from the point of unified mutational deterioration theory. Finally, an important aspect on the symmetry of the code construction, namely, the Yin-Yang duality in the genetic code will be investigated in the last section.

## 1. Mutational deterioration theory on the redundancy distribution in the genetic code

The constancy of the genetic code among different organisms is one of the most striking, interesting, and challenging phenomena in life. The mathematical relation behind the constancy intrigued many biologists and physicists [di Giulio, 1997; Trifonov et al, 1997; Freeland et al, 1998; Maeshiro et al, 1998; Judson et al, 1999; Jimenez-Montano, 1999; Knight et al, 1999; Knight et al, 2000; Freeland et al, 2000; Weberndorfer et al, 2003; Chechetkin, 2003; Copley et al, 2005; Yang, 2005; Goodarzi et al, 2005; Chechetkin, 2006]. Historically, there are two different kinds of theories regarding the origin and evolution of the genetic code [Yockey, 1992; Freeland et al, 2003]. The first approach originated from Gamow [1954]. His "Diamond code" model opened up a way to explain the origin of the universal amino acid code through the stereochemical interactions between codons or anticodons and amino acids [Woese et al, 1966; Woese, 1967; and recently, Knight et al, 2001; Yarus, 2000]. The second approach is called "frozen accident" theory. The term "frozen accident", used firstly by Crick, means that all living organisms evolved from an ancient single ancestor, and after the evolutionary expansion of the descendants started, changes in the amino acid assignments of codons were not possible [Crick, 1968]. In fact, the two theories can explain part of observations and experiments from their own standpoints, but they are by no means comprehensive. For example, the point that the canonical code is superior due to some specific fit or affinity between each amino acid and its codon (through t-RNA molecule as an adaptor)[Crick et al, 1961] has never been proved rigorously. The deviant codon assignments discovered since 1979 demonstrate that other different codes are also possible [Jukes & Osawa, 1991]. On the other hand, the formation of the genetic code could not be explained as a fully accidental event. The hydrophobic order of amino acids consistent with that of their anti-codonic dinucleotide is an important fact [Larcey et al, 1983; 1992], which shows that the codon assignments may be required thermodynamically and some stereochemical relations may exist between the amino acids and the codons. So, the historical accident and the stereo-chemical constraint both exist and play their roles together in the formation of prevalent code. In the later decades many new variants of above two theories were proposed. For example, the frozen accident is related to the amino acid alphabet expanding. Trifonov et al studied the temporal order of 20 kinds of amino acids [Trifonov et al, 1997; Trifonov, 2004]. Wong proposed the coevolution theory which indicated the coevolution existed between amino acids and codes [Wong, 1975; 1988; 2005].

From 1988 on we have proposed an alternative approach to the problem [Luo, 1988; 1989]. We started from the observation of the pattern of codon degeneracy and the investigation of the synonym redundancy distribution in the code. From the genetic code (Figure 1) we find that each degenerate codon doublet is located on the upper side or lower side of one of the 4×4 blocks in the table, that is, their first two nucleotides are same and the third ones are related by a transitional mutation, a mutation not changing the purine or pyrimidine-type of the nucleotide. We also find each degenerate codon quartet located in one of the 4×4 blocks, namely, their first two nucleotides are common in the quartet. The hexamerous multiplets all occupy one and half block. Ile and terminators both are triplet but their codons are arranged differently in the code table. These rules holds even for deviant codes. How to explain these rules? Our theory is based on following assumptions:

1. The mutation of code word （codon）causes wrong coding for amino acid or terminator. It is lethal and will be eliminated by evolution (selection). Regardless of the complexity existed in the mutation and translation mechanism and the possible alterations in the tRNAs, what we concern is only the code relation between code words and encoded amino acids. The prevalent code is a product of long-term evolution. The high universality of the code and its degeneracy rule over a wide range of organisms indicates its selective non-lethality. That is, as compared with other ideal codes, the real code is the most advantageous due to selection. Mathematically, for each ideal codon multiplet, one can define a mutational deterioration (MD) function that represents the mutational frequency of the multiplet and the deterioration caused by the mutation. The degeneracy rule of the code can be deduced from the minimization of MD function.

2. The MDs are classified into three categories and parameterized as follows: the non-synonymous transitional MD (MD caused by transitional mutation, U↔C G↔A, between non-synonymous codons), denoted by $u$, the non-synonymous transversional MD (MD caused by transversional mutation, U↔A, U↔G, C↔G, C↔A, between non-synonymous codons), denoted by $v$, and wobble MD $w_{u,}, w_v$ which describe the additional effect of the third letter mutation in a sense codon [Crick, 1966]. That is, we set

$$u_1 = u_2 = u, \qquad v_1 = v_2 = v, \qquad u_3 = u + w_u, \qquad v_3 = v + w_v$$

$$(u, v, w_u, w_v > 0)$$

for sense codons (the subscript 1, 2, or 3 means the position in a codon). For nonsense codons (terminators) $w_u = w_v = 0$ should be taken. Here we emphasize only the non-synonymous transitional and transversional mutations are considered since the synonymous mutation has no lethal effect. The MD function for an ideal multiplet is equal to the sum of mutational deterioration of all single-base mutations for codons belonging to the multiplet. The double- and triple-base mutations are neglected due to their frequencies much smaller than the single-base ones.

From these assumptions we can deduce all degeneracy rules in the genetic code [Luo,1989; Luo, 2000; Luo et al, 2002b]. Two codons that can not be related with each other through a single-base mutation (for example, UGC and CAU) are called non-neighboring. Oppositely, two codons related with each other through a single-base mutation are called neighboring. If they are related by a transitional mutation between first bases the two codons are called T1-neighboring. If they are related by a transversional mutation between first bases the two codons are called V1-neighboring. Likewise, one can define T2, T3, V2, and V3-neighboring of two codons.

For example, consider an MD comparison for all possible ideal degenerate doublets. For an ideal codon arrangement it is easy to deduce the MD function as the sum of contributions from all possible single-base mutations in the doublet. The results are,

Non-neighboring

$$\begin{aligned} MD(2) &= c_2 = 2c_1 \\ &= 6u + 12v + 2w_u + 4w_v \end{aligned} \tag{1.1}$$

T1 or T2 neighboring $$MD(2)=c_2-2u \quad (1.2)$$

V1 or V2 neighboring $$MD(2)=c_2-2v \quad (1.3)$$

T3 neighboring $$MD(2)=c_2-2u-2w_u \quad (1.4)$$

V3 neighboring $$MD(2)=c_2-2v-2w_v \quad (1.5)$$

As $u>v$, and $w_u> w_v> 0$, Eq.(1.4) yields the smallest value. The relations $u>v$ and $w_u> w_v> 0$ mean that the rate of transitional mutation is larger than transversional mutation and the rate of mutation at third position of a codon is larger than at other positions. Under these conditions, the minimization of MD would lead to the degenerate doublet taking the form of T3 neighboring. In fact, the nine amino acids of degenerate doublet in the standard code table all take this form of codon arrangement. The physics behind the above deduction is that the deterioration of nucleotide mutation comes from the amino acid substitution, and the amino acid substitution in an organism would generally lead to an amount of selective death. However, the synonymous mutation has no lethal effect. So, to reduce the mutational deterioration at best, the nucleotides in a multiplet should be so arranged that a large portion of base mutations, especially for transitional mutation in the third codon position, belong to synonymous mutations within the multiplet.

The above approach can be generalized to other degenerate multiplets. For a multiplet with degenerate degree $k$, there are $k(k-1)/2$ ways of pairing of codons. The connection of each pair may be T1, T3, V1, V3 or non-neighboring. Suppose that each codon arrangement is called a graph. There are generally $5^{k(k-1)/2}$ graphs for the multiplet with degenerate degree $k$. For a given graph, suppose there are $n_1$ connections being T1 (or T2) neighboring, $n_3$ connections being T3 neighboring, $m_1$ connections being V1 (or V2) neighboring, $m_3$ connections being V3 neighboring, and others non-neighboring. The MD for this graph is

$$MD(k)=c_k-n_1(2u)-n_3(2u+2w_u)-m_1(2v)-m_3(2v+2w_v)$$

$$c_k = k\,c_1 = k\,(3u+6v+w_u+2w_v) \quad (1.6)$$

There are 125 graphs for a degenerate triplet. Many of them are forbidden due to inconsistent connections. The parameters $n_1$, $n_3$, $m_1$, and $m_3$ for allowable graphs are listed in Table 1-1. The first line in Table gives the name $k.p.l$ for each graph ($k$ = degenerate degree, here $k$=3; $p= n_1+n_3 +m_1+m_3$ ; $l$ denotes the number of graph for given $k$ and $p$), the second line gives an example for the graph. From Table 1-1 we find that the graph 3.3.1 has minimal MD when $u>v$, $w_u> w_v> 0$ , $2w_v > u-2v$. The corresponding minimal MD level is

$$MD(3)=c_3-2u-4v-2w_u-4w_v \quad (1.7)$$

So, the degenerate triplet of codons should take this arrangement with MD (1.7). In fact, for Ile, three codons are distributed in this manner in the code table. However, for three terminators, though the MD for ideal codon arrangement is still expressed by (1.6) (with $k$=3), the relations $w_u = w_v = 0$ should be taken into account. For a graph corresponding to third column of Table 1-1 (graph 3.2.1), one has

$$MD(\mathrm{TC}) = 5u+18v \quad (1.8)$$

It is easily shown that this takes on a minimum when $u>2v$. So terminators should be arranged in this form, which is different from amino acid Ile.

For quartet there are $5^6$ graphs. The parameters $n_1$, $n_3$, $m_1$, and $m_3$ for allowable graphs are listed in Table 1-2. To save space, only graphs with $p>3$ are listed. From Tab 1-2, we find that the graph 4.6.1 has minimal MD when $u>v$, $w_u > w_v > 0$, $2w_v > u-2v$. Its MD level is

$$MD(4) = c_4 - 4u - 8v - 4w_u - 8w_v \tag{1.9}$$

So, the degenerate quartet should take this arrangement with MD given by (1.9). In fact, there are five amino acids of degenerate quartet in the standard code table that all take this arrangement of codons.

For hexamerous multiplets there are $5^{15}$ graphs. The parameters $n_1$, $n_3$, $m_1$, and $m_3$ for allowable graphs are listed in Table 1-3. To save space, only graphs with $p>6$ are listed. The graph of hexamerous multiplet can be deduced from some quartet as nucleus. Evidently, the nucleus for a graph is not unique. To give an intuitive picture, a possible nucleus for each graph is shown in the second line of Table 1-3. From Table 1-3 we find the graph 6.9.1 has minimal MD level, graph 6.9.2 – the first excited level, graphs 6.7.1 – the second excited level and graphs 6.9.3 – the third excited level when

$$u>v,\quad w_u > w_v > 0,\quad w_v > u-v,\quad w_u - w_v > u+v.$$

are assumed. The three lowest levels correspond to Leu, Arg, and Ser, respectively.

To summarize, under the assumption

$$\begin{aligned} &u > 2v, \\ &w_v > u - v, \\ &w_u - w_v > u + v \end{aligned} \tag{1.10}$$

one can deduce all degeneracy rules in the genetic code. The condition (1.10) can easily be understood since it indicates the difference between transitional and transversional mutations and the importance of wobble's mutation. From experimental data on single-base mutation in pseudo genes, one finds the rate of transitional mutation larger than transversional by a factor 2 to 3. Likewise, from comparison of rates of synonymous and non-synonymous substitution, one finds the mutational rate at the third codon position is larger than first two positions by a factor 4 to 8 [Li, 1997]. Eq (1.10) is consistent with these data. Thus, we have succeeded in deducing the codon arrangements for all amino acid and terminator multiplets with different degenerate degrees from a unified point of view. Taking the experimental data on base mutation into account and assuming $w_u / w_v = u / v$, we shall choose

$$u = 2.2v,\qquad w_u = 8.1v,\qquad w_v = 3.7v \tag{1.11}$$

in the following calculation which is in accordance with Eqs. (1.10).

Set minimum MD of the multiplet with degenerate degree $i$ denoted by $m(i)$, and set the difference (gap) between minimum MD (ground state) and first higher MD (first excited state, corresponding to some ideal codon arrangement) denoted by $\Delta(i)$. The calculation results are summarized as follows:

$$m(1) = 3u + 6v + w_u + 2w_v$$

$$m(2) = 4u + 12v + 4w_v, \qquad \Delta(2) = 2(w_u - w_v + u - v)$$

$m(3) = 7u + 14v + w_u + 2w_v$, $\Delta(3) = 2(2w_v - u + 2v)$

$m(4) = 8u + 16v$, $\Delta(4) = 4(2w_v - u + 2v)$

$m(5) = 9u + 22v + w_u + 2w_v$, $\Delta(5) = 4(2w_v - u + 2v)$

$m(6) = m(6, Leu) = 8u + 28v + 4w_v$,

$m(6, Arg) = 12u + 24v + 4w_v$, $m(6, Ser) = 12u + 28v + 4w_v$,

$\Delta(6) = 2(u - v) + 2(w_u - w_v)$ (gap between graph 6.9.3 and 6.9.1)

$$m(7) = 9u + 30v + w_u + 2w_v$$
$$m(8) = 8u + 32v \quad (1.12)$$

$$m(1, \mathrm{Ter}) = 3u + 6v$$
$$m(2, \mathrm{Ter}) = 4u + 12v$$

$$\Delta(2, \mathrm{Ter}) = 2(u - v)$$

$$m(3, \mathrm{Ter}) = 5u + 18v$$

$$\Delta(3, \mathrm{Ter}) = 2(u - 2v) \quad (1.13)$$

Simultaneously, we have

| $\frac{\Delta(2)}{m(2)}$ | $\frac{\Delta(3)}{m(3)}$ | $\frac{\Delta(4)}{m(4)}$ | $\frac{\Delta(5)}{m(5)}$ | $\frac{\Delta(6)}{m(6)}$ | $\frac{\Delta(6, Arg)}{m(6)}$ | $\frac{\Delta(6, Ser)}{m(6)}$ |
|---|---|---|---|---|---|---|
| 32% | 32% | 86% | 50% | 19% | 8% | 15% |

Note: $\Delta(6, Arg) = m(6, Arg) - m(6)$ , $\Delta(6, Ser) = m(6, Ser) - m(6)$ , and $\Delta(6) =$ gap between graph 6.9.3 and *m*(6), ground state graph 6.9.1, where *m*(6)≡*m*(6, *Leu*), *m*(6, *Arg*) and *m*(6, *Ser*) are the MD for Leu, Arg and Ser, respectively. So, as seen from the table, $\Delta(i)$ is generally not a small quantity as compared with $m(i)$. The only exceptions are $\Delta(6, Arg)$ and $\Delta(6, Ser)$. It shows that, with the exception of hexamerous degenerate codons, other MD ground states (states with minimum MD) are all relatively stable under selection. It seems difficult to attain MD-excited states through statistical fluctuation for these multiplets. However, for hexamerous degenerate multiplets, the MD gap between Arg (or Ser) and Leu is smaller. It may give a clue to understand why the hexamerous degenerate codons have taken the arrangement of Arg and Ser in the code table, which is different from the ground state Leu .

**Table 1 -1　Parameters $n_1$, $n_3$, $m_1$, and $m_3$ for allowable graphs of degenerate triplets**

| | **3.3.1** | **3.3.2** | **3.2.1** | **3.2.2** | **3.2.3** | **3.2.4** | **3.1.1** | **3.1.2** | **3.1.3** | **3.1.4** | **3.0.1** |
|---|---|---|---|---|---|---|---|---|---|---|---|
| example | AUU<br>AUC<br>AUA | AUU<br>GUU<br>CUU | AUU<br>AUC<br>GUU | AUU<br>AUC<br>CUU | AUU<br>AUA<br>GUU | AUU<br>AUA<br>CUU | AUU<br>AUC<br>GCU | AUU<br>AUA<br>GCU | AUU<br>GUU<br>GCA | AUU<br>CUU<br>GCA | AUU<br>CCU<br>GCA |
| $n_1$ | 0 | 1 | 1 | 0 | 1 | 0 | 0 | 0 | 1 | 0 | 0 |
| $n_3$ | 1 | 0 | 1 | 1 | 0 | 0 | 1 | 0 | 0 | 0 | 0 |
| $m_1$ | 0 | 2 | 0 | 1 | 0 | 1 | 0 | 0 | 0 | 1 | 0 |
| $m_3$ | 2 | 0 | 0 | 0 | 1 | 1 | 0 | 1 | 0 | 0 | 0 |

**Table 1 -2　Parameters $n_1$, $n_3$, $m_1$, and $m_3$ for allowable graphs ($p$>3 )of degenerate quartets**

| | **4.6.1** | **4.6.2** | **4.4.1** | **4.4.2** | **4.4.3** | **4.4.4** | **4.4.5** | **4.4.6** | **4.4.7** | **4.4.8** |
|---|---|---|---|---|---|---|---|---|---|---|
| example | GUU<br>GUC<br>GUA<br>GUG | GUU<br>AUU<br>CUU<br>UUU | GUU<br>GUC<br>AUU<br>AUC | GUU<br>GUC<br>CUU<br>CUC | GUU<br>GUA<br>AUU<br>AUA | GUU<br>GUA<br>CUU<br>CUA | GUU<br>GUC<br>GUA<br>AUU | GUU<br>GUC<br>GUA<br>CUU | GUU<br>AUU<br>CUU<br>GUC | GUU<br>AUU<br>CUU<br>GUA |
| $n_1$ | 0 | 2 | 2 | 0 | 2 | 0 | 1 | 0 | 1 | 1 |
| $n_3$ | 2 | 0 | 2 | 2 | 0 | 0 | 1 | 1 | 1 | 0 |
| $m_1$ | 0 | 4 | 0 | 2 | 0 | 2 | 0 | 1 | 2 | 2 |
| $m_3$ | 4 | 0 | 0 | 0 | 2 | 2 | 2 | 2 | 0 | 1 |

**Table 1 -3　Parameters $n_1$, $n_3$, $m_1$, and $m_3$ for allowable graphs ($p$>6 )of hexamerous multiplets**

| | **6.9.1** | **6.9.2** | **6.9.3** | **6.9.4** | **6.9.5** | **6.9.6** | **6.9.7** | **6.9.8** | **6.9.9** | **6.9.10** | **6.9.11** | **6.9.12** | **6.9.13** | **6.9.14** |
|---|---|---|---|---|---|---|---|---|---|---|---|---|---|---|
| nucleus | 4.6.1 | 4.6.1 | 4.6.1 | 4.6.1 | 4.6.2 | 4.6.2 | 4.6.2 | 4.6.2 | 4.4.1 | 4.4.1 | 4.4.2 | 4.4.3 | 4.6.1 | 4.6.2 |
| $n_1$ | 2 | 0 | 2 | 0 | 3 | 2 | 3 | 2 | 3 | 2 | 0 | 2 | 1 | 2 |
| $n_3$ | 3 | 3 | 2 | 2 | 2 | 2 | 0 | 0 | 2 | 3 | 2 | 0 | 2 | 1 |
| $m_1$ | 0 | 2 | 0 | 2 | 4 | 5 | 4 | 5 | 0 | 4 | 3 | 4 | 2 | 4 |
| $m_3$ | 4 | 4 | 5 | 5 | 0 | 0 | 2 | 2 | 4 | 0 | 4 | 3 | 4 | 2 |

**Table 1 -3 Parameters $n_1$, $n_3$, $m_1$, and $m_3$ for allowable graphs ($p$>6 )of hexamerous multiplets** (continued)

| | **6.8.1** | **6.8.2** | **6.8.3** | **6.8.4** | **6.8.5** | **6.8.6** | **6.8.7** | **6.8.8** | **6.8.9** | **6.8.10** |
|---|---|---|---|---|---|---|---|---|---|---|
| neucleus | 4.6.1 | 4.6.1 | 4.6.2 | 4.6.2 | 4.4.1 | 4.4.2 | 4.4.3 | 4.4.4 | 4.6.1 | 4.4.1 |
| $n_1$ | 1 | 0 | 2 | 2 | 2 | 1 | 2 | 1 | 2 | 2 |
| $n_3$ | 2 | 2 | 1 | 0 | 2 | 2 | 1 | 1 | 2 | 2 |
| $m_1$ | 1 | 2 | 4 | 4 | 2 | 3 | 2 | 3 | 0 | 4 |
| $m_3$ | 4 | 4 | 1 | 2 | 2 | 2 | 3 | 3 | 4 | 0 |

| | **6.7.1** | **6.7.2** | **6.7.3** | **6.7.4** | **6.7.5** | **6.7.6** | **6.7.7** | **6.7.8** | **6.7.9** | **6.7.10** |
|---|---|---|---|---|---|---|---|---|---|---|
| neucleus | 4.6.1 | 4.6.1 | 4.6.1 | 4.6.1 | 4.6.2 | 4.6.2 | 4.6.2 | 4.6.2 | 4.4.1 | 4.4.1 |
| $n_1$ | 0 | 0 | 1 | 0 | 2 | 2 | 3 | 2 | 2 | 2 |
| $n_3$ | 3 | 2 | 2 | 2 | 1 | 0 | 0 | 0 | 2 | 2 |
| $m_1$ | 0 | 0 | 0 | 1 | 4 | 4 | 4 | 5 | 1 | 2 |
| $m_3$ | 4 | 5 | 4 | 4 | 0 | 1 | 0 | 0 | 2 | 1 |

| | **6.7.11** | **6.7.12** | **6.7.13** | **6.7.14** | **6.7.15** | **6.7.16** | **6.7.17** | **6.7.18** | **6.7.19** | **6.7.20** |
|---|---|---|---|---|---|---|---|---|---|---|
| neucleus | 4.4.2 | 4.4.2 | 4.4.2 | 4.4.3 | 4.4.3 | 4.4.3 | 4.4.4 | 4.4.4 | 4.4.4 | 4.4.4 |
| $n_1$ | 0 | 1 | 1 | 2 | 2 | 2 | 1 | 1 | 0 | 1 |
| $n_3$ | 2 | 2 | 2 | 0 | 1 | 1 | 1 | 1 | 1 | 0 |
| $m_1$ | 3 | 2 | 3 | 2 | 2 | 1 | 2 | 3 | 3 | 3 |
| $m_3$ | 2 | 2 | 1 | 3 | 2 | 3 | 3 | 2 | 3 | 3 |

In conclusion, we have demonstrated that the redundancy distribution in the genetic code is determined by the mutational parameters, the relative rates between transitional and transversional mutations and between 1-2 codon position and 3rd codon position substitutions, and the distribution is robust relative to the parameter choice as soon as Eqs (1.10) are fulfilled.

**The averaged mutational deterioration** Set the MD of codons corresponding to a given amino acid divided by the multiplicity (degeneracy degree). We call the quantity as averaged mutational deterioration (AMD) of this amino acid. We have:

Trp, Met (singlet) AMD= $3u+6v+w_u+2w_v$ （1.14）

Cys, Tyr, His, Phe, Gln, Lys, Asn, Asp, Glu (doublet)

AMD= $2u+6v+2w_v$ (1.15)

Ile (triplet) AMD= $\frac{7}{3}u+\frac{14}{3}v+\frac{1}{3}w_u+\frac{2}{3}w_v$ (1.16)

Ser (hexamerous) AMD= $2u+\frac{14}{3}v+\frac{2}{3}w_v$ (1.17)

Arg (hexamerous) AMD= $2u+4v+\frac{2}{3}w_v$ (1.18)

Leu (hexamerous) AMD= $\frac{4}{3}u+\frac{14}{3}v+\frac{2}{3}w_v$ (1.19)

Pro, Thr, Gly, Val, Ala (quartet)

AMD= $2u+4v$ (.1.20)

Their differences are：

$$(1.15) - (1.14) = u+w_u$$

$$(1.16) - (1.15) = \frac{4}{3}v-\frac{1}{3}u+\frac{4}{3}w_v-\frac{1}{3}w_u$$

$$(1.17) - (1.16) = \frac{1}{3}u+\frac{1}{3}w_u$$

$$(1.18) - (1.17) = \frac{2}{3}v$$

$$(.1.19) - (1.18) = \frac{2}{3}u-\frac{2}{3}v$$

$$(1.20) - (1.19) = \frac{2}{3}v - \frac{2}{3}u + \frac{2}{3}w_v$$

As

$$u > v, \quad w_v > u - v, \qquad v + w_v > \frac{1}{4}(u + w_u) \tag{1.21}$$

these differences are all positive, namely, these equations (from Eq (1.14) to Eq (1.20)) are arranged in the order of decreasing AMD. In fact, from Eq. (1.10) and

$$u < 4v \ , \qquad w_u < 4w_v \tag{1.22}$$

we obtain (1.21). The parameter choice (1.11) satisfies these constraints. So, the equations from Eq (1.14) to Eq (1.20) are indeed arranged in the order of decreasing AMD. On the other hand, using the sequence data of hemoglobins a table of mutual replaceabilities of amino acids can be obtained. On the basis of this table the degree of irreplaceability of amino acid residues was established [Vokenstein, 1982]. The results are

**Table 1 -4 The relative irreplaceability of amino acid residues**

| | | | | | | | | | |
|---|---|---|---|---|---|---|---|---|---|
| Trp | Met | Cys | Tyr | His | Phe | Gln | Lys | Asn | Asp |
| 1.82 | 1,25 | 1.12 | 0.98 | 0.94 | 0.86 | 0.86 | 0.81 | 0.79 | 0.77 |
| Glu | Ile | Ser | Pro | Arg | Leu | Thr | Gly | Val | Ala |
| 0.76 | 0.65 | 0.64 | 0.61 | 0.60 | 0.58 | 0.56 | 0.56 | 0.54 | 0.52 |

By comparison of the irreplaceability and AMD we find they agree well with each other. (The only exception is Pro). The agreement means that the rarer the substitution by other residues, the higher the value of AMD. It is not surprising. Because the higher mutational deterioration of some amino acid means its larger average distance to other amino acids and in turn, its larger irreplaceability.

## 2. Adaptive minimization of global mutational deterioration and the accuracy of the genetic code

### General formalism of global mutational deterioration and hydrophilic-hydrophobic domain in the genetic code

We have derived the minimum arrangement of codons in each degenerate multiplet. This is the local minimum. Then, what is the global minimum of mutational deterioration for the code table as a whole? Is the mutational deterioration of prevalent code is globally minimized? This is a problem of the distribution of twenty kinds of amino acids on the code table. To have a clear understanding we shall investigate the block distribution of amino acids in the code, namely, the hydrophobic-hydrophilic domain-like distribution at first.

There are several methods to obtain the experimental scale of hydrophobicity. Usually they can be classified into two categories. The first is to measure the solubility difference between water and some apolar solvent. The second is to measure the tendency of an amino acid residue to be sequestered inside the folded molecule. From the theoretical standpoint, the hydrophobicity is related to all kinds of quantum interactions between solute and solvent as well as the entropy factor. But, in the final analysis, it is determined by the chemical structure of the residue. We suggest that the hydrophobicity of an amino acid is determined by the type of atoms on the end of side-chain. If the atom on the end of side-chain is NH or OH, then the amino acid is hydrophilic; If the atom on the end of side-chain is CH or SH, then the amino acid is hydrophobic. If the end is a ring, then it is hydrophilic when there exsits NH or OH in the ring, or hydrophobic otherwise. The case of Gly is more complex. The recognition site of anti-codon is NH in the end of peptide [Davydov, 1989]. Thus, we obtain the hydrophobicity scale as follows:

Hydrophilic: Arg , Lys, Asp, Glu (charged); Asn, Gln, His (strong polar); Tyr, Ser, Thr (polar); Gly;

Hydrophobic: Ile, Val, Leu, Phe (strong hydrophobic); Met, Ala, Trp, Cys (hydrophobic); Pro.

Sometimes, cysteine is classified as an independent subclass since this residue has some special properties, for instance, its ability to form disulfide bridges that plays an important role in protein folding. The above hydrophilic-hydrophobic classification of twenty amino acids is consistent with other works apart from a little difference. In Kyte and Doolittle's scale [Kyte & Doolittle, 1982] and Eisenberg's scale [Eisenberg & McLachlan, 1986] Gly is hydrophobic and Pro is hydrophilic. However, from the consideration of free energy difference, Pro has high hydrophobicity while Gly has strong hydrophilicity [Nozaki & Tanford, 1971], which is consistent with our classification based on the chemical structure.

According to the above classification we can divide Figure 1 of the genetic code into two regions. The amino acids inside the solid line are hydrophilic but outside it hydrophobic. The case for dinucleotide UC (framed by dotted line) should be considered carefully, since serine is hydrophilic but the 3'-dinucleotides in anticodon corresponding to UC is hydrophobic [Wong, 1988]. We named the above-mentioned distribution of amino acids as hydrophobic(H)-hydrophilic(P) domain [Luo, 1989]. Although the measure of hydrophobicity is not unique in biology and the amino acids with medium hydrophobicity can change their positions in hydrophobic order, one can always divide amino acids into a hydrophobic domain and a hydrophilic domain on the code table.

If the conventional base order UCAG has been changed to UCGA on the code table then the hydrophobic-hydrophilic domain can be displayed in a more symmetric fashion as seen in Figure 6 [Luo, 1992]. The meaning of the base order UCGA will be discussed in section 4.

How to explain the hydrophobic-hydrophilic domain of amino acid distribution in the code table? We shall deduce it from the global minimization of mutational deterioration of the genetic code. In the previous discussions on mutational deterioration only the difference between synonymous and non-synonymous mutations is taken into account but the more detailed differences of deterioration among amino acids in the non-synonymous mutation have not been considered. In fact, the selective death caused by amino acid replacement is an important factor of mutational deterioration which should be studied carefully. For example, if an amino acid substitution due to base mutation changes the hydrophobicity drastically, then the mutation is explicitly lethal. On the contrary, if the amino acid substitution does not alter the hydrophobicity then the lethal effect is small.

To take the difference in amino acid substitution into account we define the global MD

(GMD) (for an ideal code table $U$ ) as follows [Luo, 2000; Luo et al, 2002a; 2002b]

$$Q(U) = \sum_{i \neq j, \alpha, \beta} U_{i\alpha} U_{j\beta} f_{ij} D_{\alpha\beta} \qquad (2.1)$$

GMD $Q(U)$ is a quantity to measure the accuracy of the table $U$. Set $[U]$ to be a 64×21 matrix that represents an ideal code. $U_{i\alpha}=1$ shows the $i$-th codon coding for the α-th amino acid (or terminators); otherwise $U_{i\alpha}=0$ . In other words, $U_{i\alpha}$' s ( $i$=1…64) describe the codon distribution of the α-th amino acid (or terminator) in code. So one has

$$\sum_{i}^{64} U_{i\alpha} = \text{degeneracy degree of amino acid } \alpha$$

$$\sum_{\alpha}^{21} U_{i\alpha} = 1 \qquad (2.2)$$

$f_{ij}$ denotes the mutational deterioration for codon $i$ mutated to codon $j$ . It has been parameterized through $u$, $v$, $w_u$ and $w_v$ introduced in the previous section. Namely, if $i$ and $j$ are related by a non-synonymous single-base mutation, one has

$$f_{ij} = u. \quad \text{if } i \text{ and } j \text{ T1 or T2 – neighboring}$$

$$f_{ij} = v. \quad \text{if } i \text{ and } j \text{ V1 or V2 – neighboring}$$

$$f_{ij} = u + w_u \quad \text{if } i \text{ and } j \text{ T3 – neighboring}$$

$$f_{ij} = v + w_v \quad \text{if } i \text{ and } j \text{ V3 – neighboring} \qquad (2.3)$$

($f_{ij}$ =0 if $i$ and $j$ cannot be related by any single-base mutation). For the mutation between a terminator and an amino acid, $w_u = w_v = 0$ should be taken in Eq.(2.3). The GMD of a code table is the sum of MDs from all pairs of codons. As shown in Eq (2.1), GMD depends on two factors — one is the mutational rate and the other is the selective force, represented by the distance $D_{\alpha\beta}$ between amino acids of initial and final states. If the distance between a pair of amino acids is large (the similarity of the two is small) then the corresponding mutational deterioration will be serious. On the contrary, the small distance for a pair of amino acids suggests a weak deterioration in their replacement. So, the mutational deterioration of a code depends not only on $f_{ij}$ , but also on $D_{\alpha\beta}$—the distance between amino acids $\alpha$ and $\beta$. The common approach to define amino acid distance is based on evolutionary data (PAM matrix data). There are many new developments and applications in recent years (for example, see Wyckoff et al, 2000). However, the evolutionary approach is pure empirical and has been criticized as tautologous in its application for the study of the genetic code origin [Di Giulio, 2001]. From the standpoint of basic research we prefer using the difference of physico-chemical property between a pair of amino acids to define their distance. Following Grantham [Grantham,1974] we define the physico-chemical distance between amino acids $\alpha$ and $\beta$ as

$$D_{\alpha\beta} = \left[ c_1 \left( c_\alpha - c_\beta \right)^2 + c_2 \left( P_\alpha - P_\beta \right)^2 + c_3 \left( v_\alpha - v_\beta \right)^2 \right]^{\frac{1}{2}}$$

(here $c$ = composition, $p$ = polarity and $v$ = molecular volume). Evidently, it leads to $D_{\alpha\beta} = 0$ for $\alpha = \beta$. On the other hand, we assume $D_{\alpha,ter} = D_{ter,\alpha} =$ a large enough number ( $ter$ means terminators) due to the similarity between any amino acid $\alpha$ and terminators being very small.

The genetic code table (Fig 1) is constructed from 4×4 blocks and each block is labeled by a pair of numbers ($m$,$n$) ($m$, $n$ = 1…4, representing U,C,A,G respectively; $m$ referring to the first letter of a codon and $n$ its second letter). The GMD $Q(U)$ has the following symmetries: 1) $Q$ remains invariant when the 4×4 table is transposed, namely, the ($m$,$n$) element exchanged with ($n$,$m$) element. 2) $Q$ remains invariant when 1st and 2nd rows are exchanged with 3rd and 4th rows , or 1st and 2nd columns exchanged with 3rd and 4th columns; $Q$ remains invariant when 1st and 2nd rows (columns) are exchanged between themselves, or 3rd and 4th rows (columns) are exchanged between themselves. 3) $Q$ remains invariant when 1st and 2nd element in each block are exchanged with 3rd and 4th element; $Q$ remains invariant when 1st and 2nd element, or 3rd and 4th element in each block are exchanged between themselves. In accordance with the above symmetries the ideal code table can be classified into many representations. A representation can be identified through fixation of terminators and some amino acid on given sites of the table. Two different representations are connected by a symmetrical operation. Evidently, it is enough to investigate the GMD spectrum in one particular representation.

**Table 2-1 Amino acid distance $D_{\alpha\beta}$**

| | Tyr | His | Gln | Arg | Thr | Asn | Lys | Asp | Glu | Gly | Phe | Leu | Ala | Ser | Pro | Ile | Met | Val | Cys | Trp |
|---|---|---|---|---|---|---|---|---|---|---|---|---|---|---|---|---|---|---|---|---|
| Tyr | 0 | 83 | 99 | 77 | 92 | 143 | 85 | 160 | 122 | 147 | 22 | 36 | 112 | 144 | 110 | 33 | 36 | 55 | 194 | 37 |
| His | 83 | 0 | 24 | 29 | 47 | 68 | 32 | 81 | 40 | 98 | 100 | 99 | 86 | 89 | 77 | 94 | 87 | 84 | 174 | 115 |
| Gln | 99 | 24 | 0 | 43 | 42 | 46 | 53 | 61 | 29 | 87 | 116 | 113 | 91 | 68 | 76 | 109 | 101 | 96 | 154 | 130 |
| Arg | 77 | 29 | 43 | 0 | 71 | 86 | 26 | 96 | 54 | 125 | 97 | 102 | 112 | 110 | 103 | 97 | 91 | 96 | 180 | 102 |
| Thr | 92 | 47 | 42 | 71 | 0 | 65 | 78 | 85 | 65 | 59 | 103 | 92 | 58 | 58 | 38 | 89 | 81 | 69 | 149 | 128 |
| Asn | 143 | 68 | 46 | 86 | 65 | 0 | 94 | 23 | 42 | 80 | 158 | 153 | 111 | 46 | 91 | 149 | 142 | 133 | 139 | 174 |
| Lys | 85 | 32 | 53 | 26 | 78 | 94 | 0 | 101 | 56 | 127 | 102 | 107 | 106 | 121 | 103 | 102 | 95 | 97 | 202 | 110 |
| Asp | 160 | 81 | 61 | 96 | 85 | 23 | 101 | 0 | 45 | 94 | 177 | 172 | 126 | 65 | 108 | 168 | 160 | 152 | 154 | 181 |
| Glu | 122 | 40 | 29 | 54 | 65 | 42 | 56 | 45 | 0 | 98 | 140 | 138 | 107 | 80 | 93 | 134 | 126 | 121 | 170 | 152 |
| Gly | 147 | 98 | 87 | 125 | 59 | 80 | 127 | 94 | 98 | 0 | 153 | 138 | 60 | 56 | 42 | 135 | 127 | 109 | 159 | 184 |
| Phe | 22 | 100 | 116 | 97 | 103 | 158 | 102 | 177 | 140 | 153 | 0 | 22 | 113 | 155 | 114 | 21 | 28 | 50 | 205 | 40 |
| Leu | 36 | 99 | 113 | 102 | 92 | 153 | 107 | 172 | 138 | 138 | 22 | 0 | 96 | 145 | 98 | 5 | 15 | 32 | 198 | 61 |
| Ala | 112 | 86 | 91 | 112 | 58 | 111 | 106 | 126 | 107 | 60 | 113 | 96 | 0 | 99 | 27 | 94 | 84 | 64 | 195 | 148 |
| Ser | 144 | 89 | 68 | 110 | 58 | 46 | 121 | 65 | 80 | 56 | 155 | 145 | 99 | 0 | 74 | 142 | 135 | 124 | 112 | 177 |
| Pro | 110 | 77 | 76 | 103 | 38 | 91 | 103 | 108 | 93 | 42 | 114 | 98 | 27 | 74 | 0 | 95 | 87 | 68 | 169 | 147 |
| Ile | 33 | 94 | 109 | 97 | 89 | 149 | 102 | 168 | 134 | 135 | 21 | 5 | 94 | 142 | 95 | 0 | 10 | 29 | 198 | 61 |
| Met | 36 | 87 | 101 | 91 | 81 | 142 | 95 | 160 | 126 | 127 | 28 | 15 | 84 | 135 | 87 | 10 | 0 | 21 | 196 | 67 |
| Val | 55 | 84 | 96 | 96 | 69 | 133 | 97 | 152 | 121 | 109 | 50 | 32 | 64 | 124 | 68 | 29 | 21 | 0 | 192 | 88 |
| Cys | 194 | 174 | 154 | 180 | 149 | 139 | 202 | 154 | 170 | 159 | 205 | 198 | 195 | 112 | 169 | 198 | 196 | 192 | 0 | 215 |
| Trp | 37 | 115 | 130 | 102 | 128 | 174 | 110 | 181 | 152 | 184 | 40 | 61 | 148 | 177 | 147 | 61 | 67 | 88 | 215 | 0 |

(After Grantham, 1974)

Now we shall discuss the minimization of *Q(U)*. The largest terms in Eq (2.1) are those with (α,β)=(terminator, amino acid) or (amino acid, terminator). By splitting out the leading terms one obtains

$$Q(U) = q_{ter}(U) + Q'(U) \tag{2.4}$$

$$q_{ter}(U) = 2D\sum_{ij} \sum_{\beta \neq ter} U_{i,ter} U_{j\beta} f_{ij}$$

$$Q'(U) = \sum_{ij}^{61} \sum_{\alpha\beta}^{20} U_{i\alpha} U_{j\beta} f_{ij} D_{\alpha\beta} \tag{2.5}$$

where $D = D_{\alpha,ter} = D_{ter,\alpha} =$ const., the factor 2 comes from the equal contribution of *ter* mutating to amino acid and its reverse. $q_{ter}(U)$ is the leading term. The minimization of $q_{ter}(U)$ is just like the procedure used in deducing Eq (1.8). It leads to the minimal MD described by Eq (1.8) and the corresponding optimal $U_{i\alpha}$= (codons 1 and 2, T3- neighboring ; codons 1 and 3, T1- or T2-neighboring). Evidently, the solution of optimal $U_{i\alpha}$ is not unique. The arrangement UAA, UAG, and UGA of three codons occurred in the prevalent code is one of the minimal solutions. To remove the degeneracy, we shall discuss the minimization in a particular representation where the terminators have been fixed as in the standard code .

The next step is minimization of *Q'(U)* (Eq.(2.5), assuming the terminators have been fixed. To deduce the hydrophilic - hydrophobic domain we shall investigate a simplified model. Suppose amino acids classified into several categories and neglect their differences in each category. In this approximation, one assumes

$$\begin{aligned} D_{\alpha\beta} &= \lambda \quad \text{if } \alpha, \beta \text{ in different category} \\ &= \lambda - \delta \quad \text{if } \alpha, \beta \text{ in same category but } \alpha \neq \beta \\ &= 0 \quad \text{if } \alpha = \beta \end{aligned} \tag{2.6}$$

Denote the corresponding MD function as $Q'_{\text{model}}(U)$. We have

$$Q'_{\text{model}}(U) = \sum_{\alpha}^{20} \{ \lambda \sum_{i\in\alpha, j\notin\alpha} f_{ij} - \delta \sum_{i\in\alpha, j\in\alpha 1} f_{ij} \} \tag{2.7}$$

( $\alpha 1 \neq \alpha$ , denoting different amino acids but in the same category). The minimization of the first term leads to degeneracy rules for each amino acid multiplet which has been discussed in section 1. But there are many different distributions of amino acids satisfying the same degeneracy rules. The minimization of the second term (the term proportional to δ ) would select some from all possible distributions satisfying degeneracy rules. It will lead to H-P domain. By inspection of $D_{\alpha\beta}$ data, Tab 2-1, we shall classify 20 amino acids into three categories: H-, P-, and C(cysteine)- class from the consideration of distances. The distance between amino acids in different classes is obviously larger than that in same class as seen in Tab 2-1. The three categories of amino acids occupy three domains: 7 and 1/4 H-blocks, 7 and 1/2 P-blocks and 1/2 C-block on the code table. The distribution of these H-blocks, P-blocks and C-block on the standard code table can be found in Fig 1. Any ideal assignment (*U*) of codons leads to a particular distribution of three types of blocks and gives a $Q'_{\text{model}}(U)$. They are differentiated through δ term.

For several typical H-, P-, and C- block distributions the calculated results on δ term in $Q'_{\text{model}}(U)$ were given in literatures [Luo, 1989; Luo, 2000; Luo, 2004]. By comparison of the δ term in different block distributions we found that under conditions Eq (1.10) with the complement

$$w_v > 2u - v \tag{2.8}$$

$Q'_{\text{model}}$ reaches its minimum. There exist several minimal hydrophobic-hydrophilic distributions with the same maximal δ term. The block distribution in the standard code (Fig 1) is one of the minimal distributions. Therefore, the minimization of $Q'_{\text{model}}(U)$, i.e. the minimization of GMD, can lead to the domain-like distribution of amino acids in the prevalent code. The result is understandable because the mutational deterioration of an ideal code is minimal only when the hydrophilic and hydrophobic amino acids are arranged in two separately-connected regions in good order, so that the mutations within a group (hydrophilic or hydrophobic) comprise a larger percentage of amino acid replacements since the mutation within a group contributes a smaller deterioration to the code. Note that the equation (2.8) is only a modification of the second equation of (1.10). These equations mean under a larger transitional-to-transversional ratio and a larger wobble-to-non-wobble ratio, not only the redundancy distribution but also the hydrophilic-hydrophobic distribution are robust in the code relative to the mutational parameter choice.

### Deducing the optimal code from GMD minimization

We have succeeded in deducing the hydrophobic–hydrophilic domain distribution of amino acids through a simplified model, namely, through the minimization of $Q'_{\text{model}}(U)$. Now we will search for the global minimum of mutational deterioration $Q'(U)$, Eq.(2.5). The mutational parameters Eq.(1.11) and the amino acid distances Table 2-1 will be used in the following calculation. The minimization of $Q'(U)$ can be accomplished through permutation of the rows of matrix $U$ since one permutation equivalent to one ideal code. However, there are 61! permutations so this is computationally intractable. To simplify the calculation, we assume that the degeneracy degree has been given for each amino acid and the degeneracy rule has been satisfied for each multiplet, since the constraints on mutational parameters, Eqs. (1.10) and (2.8), have been assumed and the parameter choice Eq. (1.11) satisfies these constraints. For simplicity, the codons of hexamerous degenerate amino acids are assumed to be arranged as one quartet and one doublet obeying their degeneracy rules respectively. The problem is then converted to the permutation of 20 amino acids in 4×4 blocks of the table. Furthermore, any ideal arrangement of amino acids on the table which deviates from the minimal H-P domain distribution seriously (i.e., the distribution with H-blocks and P-blocks scattered and mixed each other) should have much higher GMD and can be neglected in the minimization. Thus, the task of searching for the global minimum of $Q'(U)$ can be completed on a PC computer. Formally, the minimization can be done in the following steps.

**Step 1**. The triplet Ile should be grouped with a codon singlet in a block. The distance between Met and Ile is 10, much smaller than Trp and Ile (Trp-Ile distance 61, see Table 2-1). So Ile shares a block with initial codon Met.

**Step 2.** The codon singlet that shares a half-block with terminator UGA should be Trp, since another singlet Met has been grouped with Ile.

**Step 3**. The terminators UAA and UAG should be grouped with a codon doublet in a block. The best candidate for the doublet is Cys, since Cys has large distance with all other amino acids.

**Step 4**. The half-block Trp-ter(UGA) should be grouped with a codon doublet, too. The best candidate of the doublet is Tyr, since the amino acid which has the smallest distance with Trp is Tyr.

**Step 5**. Phe should be grouped with Leu since their distance is small and Tyr has been grouped with Trp (both the distance between Phe and Leu, and the distance between Phe and Tyr being 22, Table 2-1).

**Step 6**. The blocks ($m$, $n$)=(1,3) and (1,4) have been fixed on account of step 2 to 4. The remaining 14 blocks are divided into 7 hydrophobic, namely (1,1),(1,2),(2,1), (2,2),(3,1),(4,1) and (4,2) (called H-blocks), and 7 hydrophilic, namely (2,3),(2,4),(3,2), (3,3),(3,4),(4,3) and (4,4) (called P-blocks). The 14 blocks code for 17 amino acids, in which there are 7 doublets, namely, Asp, Glu, Asn, Lys, Gln, His, and Phe, and 3 hexamerous multiplets — Leu, Arg and Ser. A hexamerous multiplet occupies one and a half blocks. The half-blocks are denoted by Leu(2), Arg(2) and Ser(2) respectively.

**Step 7.** The case of two doublets A and B located in one block is called "doublet bundle", denoted as (A,B). The most favorable combinations of 6 doublets (except Phe in 7 doublets which has been grouped with Leu) and Arg (2) and Ser (2) are (Asp, Glu), (Asn, Ser(2)), (Gln, His), and (Lys, Arg(2)), since the sum of above four distances takes the smallest value 141. The next favorable combinations of the 8 amino acids are (Asp,Asn), (Glu, Ser(2)), (Gln, His), and (Lys, Arg(2)). The sum of their distances is 153.

**Step 8.** The hydrophobic amino acids Leu, Phe, Ile, Met and Val occupy 4 H-blocks. The hydrophilic amino acids Arg, Ser, Asp, Glu, His, Gln, Asn and Lys occupy 6 P- blocks (by use of the most favorable combinations indicated in step 7). The amino acids Gly, Thr, Pro and Ala with medium hydrophobicity can change their positions in hydrophobic order. One of the four (say, Thr) is chosen to be filled in a P-block and other three are chosen to be filled in H-blocks. Thus we have 7 H-blocks of hydrophobic amino acids and 7 P-blocks of hydrophilic amino acids. For each distribution of H blocks we permute 7 P blocks and search for the minimal distribution (distribution with minimal GMD). Then, under the fixed minimal distribution of P blocks we permute 7 H blocks and search for the new minimal distribution. Repeating the above steps, finally one obtains the self-consistent minimal solution.

**Step 9**. Based on the result obtained in step 8, taking the possible change of hydrophobic order into account we further permute four amino acids with medium hydrophobicity – Gly, Thr, Pro and Ala and find the minimal GMD.

**Step 10**. To prove the above calculation, one can make some checks. The first check is: By use of the next favorable combinations of 8 amino acids (Asp,Asn), (His,Gln), (Glu, Ser(2)), and (Lys, Arg(2)) indicated in step 7 instead of the most favorable combinations, we repeat the steps 8 and 9 and compare the result with that obtained in step 9. The second check is: We take Cys, instead of Tyr, grouped with Trp-ter(UGA) and Tyr, instead of Cys, grouped with terminators UAA and UAG.. By the above procedures we can check if the global minimum deduced in step 9 is true.

Through steps 1 to 10, setting $v$=1 in Eq.(1.11), we obtain global minimum $Q_{\min}$ = 41722 and the corresponding minimal table shown in Figure 2 [Luo and Li, 2002a].

The minimal code has following properties. 1) The GMD spectrum near the ground state (minimal code) has abundant structure. For example, an exchange between two amino acid doublets in a block in the vicinity of the ground state leads to a change of Q value (global minimum *Q*') about 10 to 30; an exchange between some quartets (namely, Gly, Ala, Thr and Pro) that have similar property and located in the lower-left site of the minimal table also causes a change of Q value about 10 to 30. These results are related to the robustness of the genetic code. 2) We find that many doublet bundles in the minimal code are same as those in the standard code. However, the important differences are: Cys / Trp-ter bundle and Tyr / ter bundle in the standard table have been changed to Tyr / Trp-ter bundle and Cys / ter bundle in minimal table, Arg doublet / Ser doublet bundle and Lys / Asn bundle in the standard table have been changed to Ser / Asn bundle and Arg / Lys bundle in the minimal table. These changes largely lower the Q value of GMD. The point can easily be estimated from the amino acid distance $D_{\alpha\beta}$ data (see Table 2-1). 3) Three hexamerous degenerate codons in minimal table all have been arranged following the same degeneracy rule of ground state, namely, T1 or T2 neighboring between their quartet and doublet components (graph 6.9.1 of Table 1-3). 4) Amino acids with similar hydrophobicity are arranged as near as possible in the minimal table. For example, the strong hydrophobic amino acids locate in the upper-left sites of the table, the strong hydrophilic amino acids locate in the lower-right sites of the table, and the medium hydrophobic-hydrophilic amino acids in the lower left sites of the table. The result is consistent with the above analysis in simplified model.

**Deducing the standard genetic code**

By use of the same parameters one calculates GMD for the standard code and obtains $Q_{\text{std}}$ = 54940. On the other hand, if a matrix [*U*] is stochastic, namely, the amino acid distribution is assumed to be random and the degeneracy rules on synonymous codon arrangement have been broken to the utmost extent, one obtains the maximal *Q* values, $Q_{\max}$ , near $1.75 \times 10^5$. By use of $Q_{\max} - Q_{\min}$ as a measure of maximum distance one finds the distance between the standard code and the minimal code (i.e. $Q_{\text{std}} - Q_{\min}$ ) about 9.86% of the maximum .

Why does Nature select the standard code rather than the minimal code to encode amino acids? The point is not difficult to understand since optimization (minimization) alone could not determine the structure of the prevalent genetic code. Not only the optimization (minimization) with respect to some parameters, but also the adaptive constraints in the early stage of evolution (the abundance of pre-synthesized amino acids, the precursor-product relations in biosynthetic pathways, etc) should be taken into account. The coevolution theory suggests that early on in the genetic code, only precursor amino acids were codified and later, as these precursors gave rise to new products, their codons underwent subdivision and some of the codons of each precursor were transferred to its product [Wong, 2005]. The optimization of GMD means the error minimization of the genetic code. The error minimization in the previous paragraph was done under the constraint of 20 amino acids with the same multiplicity distribution as in the standard code. In fact, the number of encoded amino acids and their degeneracy degrees changes in evolution. If there is enough knowledge on the amino-acid chronology (including the historical variation on the degeneracies of these amino acids) then we are able to deduce a more real picture on the genetic code evolution through GMD minimization under the varying constraints.

Trifonov (2004) indicated two important features of amino acid evolution: the amino acids synthesized in Miller experiments appeared first, and those associated with codon capture events (when all 64 triplets are already engaged and codons for new amino acid have to be captured from the established codon repertoires) came last. Due to lack of the knowledge on the amino acid degeneracy we propose a simplified model as follows. Assume GMD minimized under the same multiplicity distribution of 20 amino acids as in the prevalent standard code and introduce two additional constraints. The first constraint is (Cys, Trp) bundle and (Lys, Asn) bundle which are related to the later evolution stage of codon capture events. Lys and Asn have a common precursor Asp, while Cys and Trp have a common precursor Ser [Wong, 1988]. The precursor amino acid may have been encoded by some codons. The codons of Cys and Trp may also borrowed from original repertoire UGN for terminators. The second constraint is regarded to the early stage of amino acid evolution. We assume the initial fixation of Gly, Ala, Ser(4) (the quartet component of Ser) and Arg(4) (the quartet component of Arg) in the code, namely, Gly encoded by GG, Ala encoded by GC, Ser(4) encoded by UC and Arg(4) encoded by CG.. The meaning of this assumption is: Gly, Ala, and Ser were early amino acids [Wong, 1988; Trifonov, 2004]; Ser(4) fixed in hydrophobic region of the table should be a frozen accident; Arg was possibly recruited earlier due to its ability to interact with and stabilize nucleic acids by ionic forces [Houen, 1999] or due to its significant probability of codon/binding site association in the earlier RNA world [Knight et al, 2000]. So, these four amino acids encoded by GG, GC, UC and CG may be an earlier event. Under these two constraints, by use of the same calculation given in previous paragraph, we can deduce the standard code table through minimization of $Q'(U)$ logically. The main steps are:

**Step 1.** Cys and Trp-ter(UGA) are fixed in the block (1,4) and Tyr and ter(UAA/UAG) are fixed in the block (1,3) due to the assumption of (Cys, Trp) bundle.

**Step 2.** Met should be grouped with Ile ,and Phe should be grouped with Leu as stated in step 1 and 5 of previous paragraph. Under the assumption of (Lys, Asn) bundle the most favorable combinations of other doublets, namely Asp, Glu, Asn, Lys, Gln, His, Arg(2) and Ser(2), should be: (Lys,Asn), (Asp, Glu), (His,Gln), and (Ser(2)), Arg(2)). Considering the early fixation of Ser(4) and Arg(4) the coding of Arg(2) and Ser(2) may be a later independent event.

**Step 3.** Following our assumption, the blocks (1,2) and (4,2) has been filled in by Ser(4) and Ala respectively, the blocks (2,4) and (4,4) has been filled in by Arg(4) and Gly respectively. So, only five blocks in H- domain, namely, (1,1), (2,1), (2,2), (3,1) and (4,1), and five blocks in P- domain, namely, (2,3), (3,2), (3,3), (3,4) and (4,3), are to be determined. By permutation of these ten blocks we have succeeded in deducing the minimum of GMD and therefore proved that the standard code is $Q'(U)$ minimal under two constraints.

So, the twenty amino acids in the standard genetic code are distributed by the principle of minimization of GMD function [Luo and Li, 2002a].

In the deduction of the standard code we have obtained a table of intermediate case. If only the first constraint—doublet bundles of Cys / Trp and Lys / Asn—is introduced the minimization of $Q'(U)$ will lead to a table with $Q = 51039$. Its distance to the standard table is 2.91% of ($Q_{max} - Q_{min}$) [Luo, 2000; 2004]. So, to deduce the standard code by use of GMD minimization, the constraint on the early fixation of several amino acids (Gly, Ala, Ser and Arg) on the table as the initial condition is necessary.

The above discussions are held under mutational parameters given by Eq (1.11). However, the results are insensitive to the parameter choice. By changing mutational parameters in the range of experimental data (the transitional-to-transversional ratio about 2-3 and the synonymous-to-nonsynonymous ratio about 4-8) the standard code can always be deduced from the minimization of $Q'(U)$. Likewise, the results do not change substantially under some possible alteration of amino acid distances. For example, if the doublet bundles and other amino acid clustering rules (described in steps 1 to 7 of the minimization of GMD) remain unchanged by use of new distances and if the new distances are still classified into two or three categories according to hydrophobicity scale, then the basically same code can be deduced [Luo and Li, 2002a].

In present study the multiplicity of each amino acid (and stop codons) has been assumed in advance. In fact, a codon may disappear from a coding sequences due to some mutational pressure, and then it reappears and acquires a new function, which results in the change of multiplicity distribution of codons. If the multiplicities of some amino acids and terminators have been changed, the minimal code should be deduced by use of new multiplicity constraints. So, the deviant assignment of codons and the evolvability of the genetic code could be accounted for in a generalized mutational deterioration theory. The point will be discussed in the following section.

**Remarks**

1. We have proposed a unified theory on the construction and evolution of the genetic code—from the local MD minimization of a codon multiplet to the global MD minimization of the whole table. The theory explains the robustness of synonym redundancy distribution of codons and the hydrophilic-hydrophobic distribution of amino acids in the genetic code and these properties have been used for parameter choice and computational check in the global MD minimization. The meaning of GMD (Eq (2.1)) is twofold. On the one hand, the GMD can be regarded as a measure of non-fitness of the genetic code and its minimization is comparable with the Wright's adaptive landscape theory [Wright, 1932]. The minimization of the non-fitness through changing amino acid code reflects the real selection process in the code evolution. On the other hand, the GMD can be looked as an error function which contains two factors, base-mutational error and translational error. As compared with other error minimization of the genetic code, the two factors can be estimated independently in our theory. The mutational error can be minimized by the parameter choice based on the determination of synonym redundancy distribution. The translational error is minimized through the appropriate arrangement of amino acids on the optimized hydrophilic-hydrophobic domain. Simultaneously, different from Haig and Hurst (1991), the division of codon space (the 64 possible codons) into 21 nonoverlapping sets is not fixed in our theory but is changeable with amino acid replacement in variant ideal codes. Therefore, the minimal table, Fig 2, obtained by us is a unique one, different from those in other error minimization theory, for example, Fig 3 given by Di Giulio (1994) and Fig 4 given by Freeland and Hurst (1998).

2. In our approach, the prevalent standard code has been deduced logically from GMD minimization under some constraints. It has a lower MD value, but not the minimal one (deviating from the minimum about 9.86%). The result is reasonable due to the existence of constraints that relate to the amino acid expansion and the frozen accident occurred in the early stages. In some error minimization theory [Freeland and Hurst, 1998], it was argued that the standard code is "one in a million" event but how the prevalent standard code emerged from the evolutionary history is not clear. The merit of our approach is: we demonstrate that the natural code is not far from the minimal code and it is evolutionary accessible through introducing some constraints that reflect the adaptation in early evolution. The remarkable capacity of the proposed approach is due to GMD not only a measure of error, but also a quantity for describing the adaptive evolution of the genetic code. In the meantime, that the standard code is deducible through GMD minimization under two constraints adapted to the early environment also infers the big-bang-like formation of the standard code in a relatively short time after the Last Universal Common Ancestor of extant life (LUCA) [Knight et al, 2000; Chechetkin, 2003].

3. The mutational deterioration of molecular sequence [Ji and Luo, in Luo, 2000]. The concept of mutational deterioration for the genetic code can be generalized to molecular sequence. Set $P(j)$ the normalized frequency of codon $j$ in sequence, $\sum_j p(j) = 64$ . Define the mutational deterioration of molecular sequence

$$J = \sum U_{i\alpha} U_{j\beta} p(i) f_{ij} D_{\alpha\beta} \tag{2.9}$$

Eq (2.9) is reduced to $Q(U)$, Eq (2.1), as $P(i)$=1. The meaning of $J$ is a measure of natural selection strength on molecular sequence. Through calculation we find the differences of $J$ among various coding sequences are generally smaller than 5%. Virus, phage and Ras oncogene have comparatively large $J$, which may be related to the stronger mutation ability or selective death of these genes.

## 3. Evolution of the genetic code from the viewpoint of mutational deterioration theory

The evolution of the genetic code is closely related to the amino acid expansion and the change of the synonym multiplicity in the genetic code. In the previous section the GMD (non-fitness of the code) minimization was accomplished under given degeneracy degrees of amino acids and terminators. The ideal code (represented by $U_{i\alpha}$ in Eq (2.1)) is the coordinate of the landscape, and the encoded amino acid number and the degeneracy degree of each multiplet (constraints Eq (2.2)) determines the adaptive landscape of the code. Now we will discuss the possible change on the constraints of $U_{i\alpha}$ and therefore the alteration of the fittest genetic code. Since 1979 a number of departures or changes from the universal genetic code have been discovered in mitochondria. It was pointed out that mitochondria had very small genomes and, in contrast to whole organisms, can tolerate changes in the code. However, this changed in 1985. Some deviant codon assignments have also been discovered in nuclear genome [Barrell et al, 1979; Jukes & Osawa, 1991]. The deviant assignments of codons are summarized in Table 3-1[Maeshiro & Kimura, 1998, Knight et al, 2001]. In 30 deviant assignments there are 16 cases for stop codons changing to sense codons, 2 cases for the reversed reassignment (sense codons changing to stop codons), and 12 cases related to alternative codes for amino acids. The latter includes:

① AUA (Ile) codes for Met deviantly (four cases);

② AGR(Arg) codes for Ser deviantly (three cases);
③ AGR(Arg) codes for Gly deviantly (one case);
④ AAA(Lys) codes for Asn deviantly (two cases);
⑤ CUN (Leu) codes for Thr deviantly (one case); and
⑥ CUG (Leu) codes for Ser deviantly (one case) (3.1)

**Table 3 -1 Deviant assignments of codons**

| | Codon | Standard code | Abnormal code | Representative system |
|---|---|---|---|---|
| 1a | UGA | stop | Trp | Mitochondrial yeasts |
| 1b | AUA | Ile | Met | |
| 1c | CUN | Leu | Thr | |
| 2a | UGA | stop | Trp | Mitochondrial platyhelminths |
| 2b | AAA | Lys | Asn | |
| 2c | AGR | Arg | Ser | |
| 2d | UAA | stop | Tyr | |
| 3a | UGA | stop | Trp | Mitochondrial nematoda |
| 3b | AGR | Arg | Ser | arthropoda |
| 3c | AUA | Ile | Met | mollusca |
| 4a | UGA | stop | Trp | Mitochondrial echinodermata |
| 4b | AAA | Lys | Asn | |
| 4c | AGR | Arg | Ser | |
| 5a | UGA | stop | Trp | Mitochondrial tunicata |
| 5b | AUA | Ile | Met | |
| 5c | AGR | Arg | Gly | |
| 6a | UGA | stop | Trp | Mitochondrial vertebrata |
| 6b | AUA | Ile | Met | |
| 6c | AGR | Arg | stop | |
| 7a | UGA | stop | Trp | Mitochondrial euascomycetes |
| 8a | UAG | stop | Leu | Mitochondrial |
| 8b | UAG | stop | Ala | in some green plants * |
| 8c | UCA | Ser | stop | |
| 9a | UGA | stop | Trp | Nuclear mycoplasma |
| 10a | UGA | stop | Cys | Nuclear euplotes |
| 11a | UAR | stop | Gln | Nuclear acetabularia |
| 12a | UAG | stop | Gln | Nuclear blepharisma |
| 13a | CUG | Leu | Ser | Nuclear candida |
| 14a | UGA | stop | SeCys | Nuclear** |
| 15a | UAG | stop | PyLys | Nuclear** |

(After Maeshiro et al, 1998; * 8a, 8b and 8c taken from Knight et al, 2001; ** new amino acid )

These discoveries revealed that the genetic code is still evolving. Subsequently, two evolutionary theories, codon capture theory and ambiguous intermediate theory were proposed [Knight et al, 2001; Santos et al, 2004]. In both theories the evolutionary mechanisms are complex and diverse, including base medication and RNA editing in tRNAs, genetic code ambiguity, genome base composition, codon usage and codon reassignment, etc. The alterations in the tRNA, the mutation or disappearance of some tRNA species, is a key step in the alternative code evolution. To follow each detail about tRNA alteration theoretically is not easy. Ignoring these redundant and unnecessary details we shall give a quantitative observation on the deviant codes from the mutational deterioration theory [Luo, 1989; Luo et al, 2002b]. There are five evolutionary modes on the alternative genetic codes:

*Mode* 1 Reassignment of a stop codon to a sense codon (via codon capture, Santos et al, 2004). In the mode the constraint conditions in GMD minimization should be changed from Eq (2.2), namely

$$\sum_i^{64} U_{i\alpha} = R_\alpha \text{ (degeneracy degree of multiplet } \alpha\text{)}, \quad \sum_\alpha^{21} U_{i\alpha} = 1$$

to

$$\sum_i^{64} U_{i\alpha} = R_\alpha\ (\alpha \neq \alpha_0, \tau),$$

$$\text{or} \quad = R_\tau\text{-1}\ (\alpha = \tau, \text{ terminator})$$

$$= R_{\alpha 0}\text{+1}(\alpha = \alpha_0, \text{ some amino acid})$$

$$\sum_\alpha U_{i\alpha} = 1 \tag{3.2}$$

The codon reassignment can be viewed as a virtual codon interaction described by equation

$$A_i + T_j \rightarrow A_{i+1} + T_{j\text{-}1} \tag{3.3}$$

or

$$T_j \rightarrow A_1 + T_{j\text{-}1} \qquad \text{(for case 14a,15a)} \tag{3.4}$$

where $A_i$ describes an amino acid with codon multiplicity $i$ and $T_j$ terminators with multiplicity $j$. Following Eq. (2.4) the leading term in GMD is $q_{ter.}$ One may assume $D_{\alpha,ter}(D_{ter,\alpha})$ is a large number as compared with other terms. So, the process (3.3) (3.4) will lower the mutational deterioration and is selective-favorable. This explains 16 cases for stop codons changing to sense codons in Table 3-1.

*Mode* 2 Reassignment of a sense codon to a stop codon. The constraint conditions in GMD minimization should be changed from Eq (2.2) to

$$\sum_i^{64} U_{i\alpha} = R_\alpha\ (\alpha \neq \alpha_0, \tau),$$

$$\text{or} \quad = R_{\alpha 0}\text{-1}\ (\alpha = \alpha_0)$$

$$= R_\tau\text{+1}\ (\alpha = \tau)$$

$$\sum_{\alpha} U_{i\alpha} = 1 \quad (3.5)$$

The virtual codon interaction equation reads

$$A_i + T_j \rightarrow A_{i\text{-}1} + T_{j+1} \quad (3.6)$$

The GMD increases in the process if $D_{\alpha,ter} (D_{ter,\alpha})$ is a large enough number.

*Mode* 3  Sense codon reassignment via codon capture (Santos et al, 2004; Knight et al, 2001). The constraint conditions in GMD minimization is changed from Eq (2.2) to

$$\sum_{i}^{64} U_{i\alpha} = R_\alpha (\alpha \neq \alpha_0, \beta),$$

$$\text{or} = R_{\alpha 0}\text{-}1 \ (\alpha = \alpha_0)$$

$$= R_\beta\text{+}1 \ (\alpha = \beta)$$

$$\sum_{\alpha} U_{i\alpha} = 1 \quad (3.7)$$

The codon interaction equation is of the form

$$A_i + A_j \rightarrow A_k + A_l \qquad (i\text{+}j = k\text{+}l) \quad (3.8)$$

where $i = \alpha_0$, $j = \beta$, $k = \alpha_0 - 1$, $l = \beta + 1$ for the present case.

*Mode* 4  Sense codon reassignment via ambiguous intermediate (Santos et al, 2004; Knight et al, 2001). The constraint conditions in GMD minimization are changed from Eq (2.2) to

$$\sum_{i}^{64} U_{i\alpha} = R_\alpha (\alpha \neq \alpha_0, \beta),$$

$$\text{or} = R_{\alpha 0} \ (\alpha = \alpha_0)$$

$$= R_\beta\text{+}1 \ (\alpha = \beta)$$

$$\sum_{\alpha} U_{i\alpha} = 1 \ (i \neq i_0), \quad \text{or} \quad \sum_{\alpha} U_{i\alpha} = 2 \ (i = i_0, \text{ coding for } \alpha_0 \text{ and } \beta) \quad (3.9)$$

for ambiguous intermediate and finally to

$$\sum_{i}^{64} U_{i\alpha} = R_\alpha (\alpha \neq \alpha_0, \beta),$$

$$\text{or} = R_{\alpha 0\text{-}1} \ (\alpha = \alpha_0)$$

$$= R_\beta\text{+}1 \ (\alpha = \beta)$$

$$\sum_{\alpha} U_{i\alpha} = 1 \tag{3.10}$$

The final virtual codon interaction equation is also of the form of Eq (3.8).

*Mode* 5 Reassignment via two steps. The first step is same as *Mode* 2, the reassignment of a sense codon to a stop; then the second step (as *Mode* 1) follows, the reassignment of the new stop codon to another sense codon. The successive codon interaction equations in two steps are

$$\begin{aligned} &A_{R_{\alpha 0}} + T_{R_\beta} \rightarrow A_{R_{\alpha 0}-1} + T_{R_\beta +1} \\ &T_{R_\beta +1} + A_{R_\gamma} \rightarrow T_{R_\beta} + A_{R_\gamma +1} \end{aligned} \tag{3.11}$$

and the total codon interaction is the sum of above two equations

$$A_{R_{\alpha 0}} + A_{R_\gamma} \rightarrow A_{R_{\alpha 0}-1} + A_{R_\gamma +1} \tag{3.12}$$

again in the form of Eq (3.8).

The GMD variations in *Mode* 1 and 2 have been indicated above. Now we will discuss the variation of optimal value of GMD in *Mode* 3 to 5 where the change of constraint conditions is irrespective of stop codons. As stated before, the optimal GMD can always be calculated through global minimization under given constraints. The calculation is a tedious task. However, for the reassignment only related to amino acids but no terminators (as described by Eq (3.8) or (3.12)), an approximation, called independent amino acid approximation (IAAA) can be adopted. IAAA means that all differences of amino acid distances in *Q'*(*U*) (Eq.(2.5)) have been neglected, *Q'*(*U*)=constant. In this approximation the optimal *Q'*(*U*) can easily be deduced. For given code with ideal multiplicity distribution $\{n_j\}$ (*j*=multiplicity) the approximation leads to

$$Q'(U_{\min}) = \sum_j n_j m\ (j) \tag{3.13}$$

where $U_{\min}$ means the minimal code for given multiplicity distribution $\{n_j\}$, and *m(j)* represents the corresponding local minimum of MD that has been found in Eqs (1.12). Since an alternative genetic code contains only one or a small number of deviant codon reassignments the IAAA is a good approximation. By use of Eq. (3.13) we are able to deduce the GMD variation in codon reassignment *Mode* 3 to 5 immediately.

Before the calculation of GMD variation we shall check the reliability of expression (3,13) at first. The ideal multiplicity distribution $\{n_j\}$ is supposed to satisfy the constraints

$$\begin{aligned} &\Sigma\, n_j = 20 \\ &\Sigma\, j\ n_j = 61 \end{aligned} \tag{3.14}$$

From Lagrange multiple method one has

$$\delta\Big(\sum_j n_j m(j) - \lambda \sum_j n_j - \mu \sum_j j n_j\Big) = 0 \tag{3.15}$$

It gives

$$m(j) = \lambda + \mu\ j \tag{3.16}$$

One can easily check that *m*(*j*) given by Eq (1.12) (with parameter choice (1.11)) satisfies Eq (3.16) approximately. So the code table consisting of degenerate multiplets deduced from local minima of MD is approximately globally minimized. However, no information about $\{n_j\}$ has been obtained from above deduction. There is much room for the choice of multiplet distribution in code.

The minimization of GMD under given constraints (amino acid number and multiplicity) is a process of selective optimization in the evolution. The choice from the comparison of two optimal codes under different constraints has the similar meaning of selective optimization. Such a deduced code with lower GMD should be selective-favorable. Since the codon reassignments in *Mode* 3 to 5 can be reduced to the fundamental virtual process (3.8) we define the selective potential $R$

$$R = (m(i) + m(j) - m(k) - m(l)) / (m(i) + m(j)) \tag{3.17}$$

for the process. The reassignment of codons will be selective-favorable if $R > 0$. If the linear relation (3.16) holds rigorously then $R=0$ for all processes of type (3.8). But Eq (3.16) is only an approximate one and $R$ differs from zero in reality. So the reassignment of codons may increase or decrease the fitness of the code if $R > 0$ or $< 0$ respectively [Luo, 1989; Luo et al, 2002b]

The deviant codon assignment ① of Eq (3.1) can be expressed as

$$A_1 + A_3 \rightarrow A_2 + A_2$$

and by use of Eqs (3.17) (1.12) and (1.11), one has

$$R = \frac{2u - 4v + 2w_u - 4w_v}{10u + 20v + 2w_u + 4w_v} = 2.5\% \tag{3.18}$$

which is selective-favorable. The deviant codon assignment ② is also selective-favorable, since from

$$A_6\,(\text{Arg}) + A_6\,(\text{Ser}) \rightarrow A_4 + A_8$$

we have

$$R = \frac{8u + 4v + 8w_v}{24u + 52v + 8w_v} = 38\% \tag{3.19}$$

The deviant codon assignment ③ can be explained in the same way, through

$$A_6\,(\text{Arg}) + A_4\,(\text{Gly}) \rightarrow A_4 + A_6$$

and

$$R = \frac{4u - 4v}{20u + 40v + 4w_v} = 4.9\% \tag{3.20}$$

By use of the same calculation, for the deviant codon assignment ④ represented by $A_2 + A_2 \rightarrow A_1 + A_3$ one has

$$R = \frac{-2u + 4v - 2w_u + 4w_v}{8u + 24v + 8w_v} = -2.5\% \tag{3.21}$$

For the deviant codon assignment ⑤ represented by $A_6(\text{Leu}) + A_4 \rightarrow A_2 + A_8^*$ (here star means $A_8$ not at the MD minimum) one has

$$R = \frac{-4u}{16u + 44v + 4w_v} = -9.4\% \qquad (3.22)$$

For the deviant codon assignment ⑥ represented by $A_6(\text{Leu}) + A_6(\text{Ser}) \to A_5^* + A_7^*$ one has

$$R = \frac{-4u - 4v - 2w_u - 4w_v}{20u + 56v + 8w_v} = -34\% \qquad (3.23)$$

The above deviant codon assignments can be classified into 3 categories. The reassignments of class ① ② and ③ (8 in 12 cases of Eq (3.1) lower the GMD value as compared with the standard code. The reassignments of class ④ and ⑤ (3 in 12 cases) leads to a higher GMD but near the standard code. On may assume that these reassignments ① to ⑤ follow the evolutionary *Mode* 3 and 4. So, the general trend of the genetic code evolution is towards a lower GMD (or keeping the value unchanged). However, the deviant codon assignment ⑥ (of Eq (3.1)) has R = -34%, which should be explained by other evolutionary mechanism. We assume that the reassignment of codon CUG from Leu to Ser in *Candida cylindracea* (class ⑥) follows the evolutionary *Mode* 5. Different from other four modes, the *Mode* 5 consists of two steps. Since the first step is the reassignment of a sense codon to a stop codon that makes GMD increasing, the codon reassignment class ⑥ shows uncommon character of GMD-increasing. The assumption that the alternative genetic codes of class ⑥ evolves across an intermediate stop codon should wait for further test. Of course, the tRNA Leu and tRNA Ser are structurally similar and they can be conversed to each other by changing several nucleotides. The factor is also important for understanding the reassignment.

In summary, the evolution of the alternative genetic code is classified into five categories: two related to the reassignment of a stop codon to a sense codon or its reverse and three related to the reassignment of a codon between different amino acids. The variation of optimal GMD is an important quantity for describing the evolvability of the code. In IAAA approximation it can be calculated through the local minima of MD (selective potential *R*, Eq (3.17)). From the 30 reassignments of codons (Table 3-1), we find only three cases, namely 6c 8c (referring to *Mode* 2, the reassignment of a sense codon to a stop) and 13a (referring to *Mode* 5, that includes an intermediate step of the reassignment of a sense codon to a stop) are explicitly GMD-increasing. It is interesting to note that, apart from the reassignment of a sense codon to a stop codon, the evolution of alternative genetic code has a general trend of GMD non-increasing which reflects the selection on the code. In fact, many ambiguities of the intermediate (as in the ambiguous intermediate theory) may have been cleared up naturally in the evolution through the selection role of MD minimization and the finally observed reassignment is a selective- advantageous or neutral one ($R \geq 0$). As for the abnormality of GMD variation in the reassignment of a sense codon to a stop codon, it may be partly due to the lack of an accurate calculation method on the physico-chemical distance between amino acid and terminator since we have generally assumed $D_{\alpha,ter}(D_{ter,\alpha})$ a very large constant in all cases of GMD calculation.

Synonym **multiplicity distribution in the genetic code** In the last paragraph we will give an explanation on the distribution of codon multiplicities in the genetic code .

Consider the fundamental process (3.8) and calculate the total MD for a pair of multiplets with given codon number $N = i + j = k + l$. By use of Eqs. (3.8) (1.12) and parameter choice (1.11) with *v*=1 we obtain the total MD in the ground (low-lying) and the first excited states shown in Table 3–2. From the table we find low-lying pairs (with minimal total MD) $\underline{2}+\underline{2}$ for *N*=4, $\underline{1}+\underline{4}$ for *N*=5, $\underline{2}+\underline{4}$ for *N*=6, $\underline{3}+\underline{4}$ for *N*=7, $\underline{4}+\underline{4}$ for *N*=8, $\underline{1}+\underline{8}$ for *N*=9, $\underline{2}+\underline{8}$ for *N*=10, $\underline{3}+\underline{8}$ for *N*=11, and $\underline{4}+\underline{8}$ for *N*=12, etc. No $A_5$ and $A_7$ occur in low-lying pairs. This explains the multiplicity distribution in the standard code and the disappearance of $A_5$ and $A_7$ in it. They may occur in abnormal code but scarcely. $A_6$ does not occur in low-lying pairs, too, but it occurs in the first excited pairs near the ground pairs (namely, in $\underline{4}+\underline{6}$ and $\underline{1}+\underline{6}$). The calculation also shows that the pair $\underline{2}+\underline{2}$ is slightly lower than $\underline{1} + \underline{3}$, so the doublet occurs more frequently in the code table. If the virtual 3-body interaction

$$A_i + A_j + A_k \rightarrow A_l + A_m + A_n$$

is taken into account the above conclusion remains unchanged. For example, for case N=7 the low-lying state is $\underline{1}+\underline{2}+\underline{4}$ and the first excited state is $\underline{2}+\underline{2}+\underline{3}$ which are comparable with Table 3-2.

**Table 3 -2 Total MD for a pair of amino acids with given *N***

| *N* | Low-lying | $(MD)_0$ | First excited | $\frac{(MD)_1}{(MD)_0}-1$ |
|---|---|---|---|---|
| 4 | 2+2 | 71 | 1+3 | 3% |
| 5 | 1+4 | 62 | 2+3 | 30% |
| 6 | 2+4 | 69 | 1+5 | 24% |
| 7 | 3+4 | 79 | 1+6 | 12% |
| 8 | 4+4 | 67 | 2+6 | 43% |
| 9 | 1+8 | 78 | 4+5 | 17% |
| 10 | 2+8 | 85 | 4+6 | 11% |
| 11 | 3+8 | 96 | 1+10 | 8.6% |
| 12 | 4+8 | 83 | 2+10 | 35% |

$(MD)_0$ and $(MD)_1$ mean the total MD value for a pair of amino acids in low-lying state and the first excited state respectively.

## 4. Yin-Yang duality in the genetic code

### The group-theoretic symmetry behind the genetic Code

Group theory is an appropriate tool for studying the symmetry of a system. For continuous symmetry the groups with 64 dimensional irreducible representations are *SU*(2), *SU*(3), *SU*(4), *Sp*(4), *Sp*(6), *SO*(13), *SO*(14) and $G_2$. The *Sp*(6) symmetry was introduced in the genetic code study by several authors [Hornos & Hornos, 1993]. But these continuous symmetries seem too high to describe the genetic code. Should such a high symmetry sp(6) among four nucleotides exist in code, it must be seriously broken. But the decomposition of sp(6) symmetry did not reflect the real process of temporal refinement in the codon recognition [Nieselt-Struwe & Wills, 1997]. Therefore, we shall consider the discrete symmetry. Following Cayley theorem, any group G of order *n* is isomorphic with a subgroup of the symmetric group $S_n$ [Hamermesh, 1962]. To describe the symmetry among 4 nucleotides, the group $S_4$ is most appropriate. So the triplet code should be described by $S_{12} \supset S_4 \otimes S_4 \otimes S_4$. Of course, the $S_4$ symmetry may be still too high and it should be broken further. $S_4$ contains two subgroups of order 4. One is cyclic group $Z_4$ with elements

$$Z_4: \{(1\,2\,3\,4), (1\,3)(2\,4), (1\,4\,3\,2), e\}$$

$$\text{or} \quad \{e, a, b\,(=a^2), c\,(=a^3) \mid a^4=e\} \qquad (4.1)$$

(*e* means the identity). Another is Klein- 4 group $V_4$. Its elements are

$$V_4: \{(1\,2)(3\,4), (1\,3)(2\,4), (1\,4)(2\,3), e\}$$

$$\text{or} \quad \{e, a, b, c \mid a^2 = b^2 = c^2 = e,\ ab = ba = c, \ \text{etc}\} \qquad (4.2)$$

Which is the most appropriate candidate for describing the symmetry behind the genetic code, $V_4$ or $Z_4$? The elements in $V_4$, apart from the identity, are all 2-cycles. They may have clear biological meaning. While in $Z_4$, the elements include 4-cycles, such as (1 2 3 4),(1 4 3 2), etc. which lack biological meaning. So, $V_4$ is the best candidate. The Klein 4-group as a relevant group-theoretic description has been discussed in literatures [Finley *et al*, 1982; Jimenez-Montano, 1999; Luo, 2000].

The elements of $V_4$ can be defined through

$$\hat{\alpha}(abcd) = (badc)$$
$$\hat{\beta}(abcd) = (dcba)$$
$$\hat{\gamma}(abcd) = (cdab)$$
$$\hat{e}(abcd) = (abcd) \qquad (4.3)$$

Set the relation between four nucleotides and (*a,b,c,d*) as

$$\begin{aligned} U &= a+b+c+d \\ C &= a+b-c-d \\ G &= a-b+c-d \\ A &= a-b-c+d \end{aligned} \qquad (4.4)$$

Evidently, U,C,G,A are eigenstates of $\hat{\alpha}$、$\hat{\beta}$、$\hat{\gamma}$、$\hat{e}$ respectively. Their eigenvalues are given in Table 4 -1.

**Table 4-1 The eigenvalues of $V_4$**

| | **U** | **C** | **G** | **A** |
|---|---|---|---|---|
| $\hat{e}$ | +1 | +1 | +1 | +1 |
| $\hat{\alpha}$ | +1 | +1 | –1 | –1 |
| $\hat{\beta}$ | +1 | –1 | –1 | +1 |
| $\hat{\gamma}$ | +1 | –1 | +1 | –1 |

So, $\hat{\alpha}$ is the operation classifying purine ($\hat{\alpha} = -1$) and pyrimidine ($\hat{\alpha} = +1$), $\hat{\beta}$ is the operation classifying strong bond ($\hat{\beta} = -1$) and weak bond ($\hat{\beta} = +1$). However, because there is sharp distinction in physicochemical properties between different purines, pyrimidines, strong bonds or weak bonds, the $V_4$ symmetry should be broken further. The broken $V_4$ symmetry can be manifested through Yin-Yang duality (see below).

**The Duality of Genetic Code**

According to Chinese traditional medicine and ancient philosophy, life is the unity of a pair of contradictory factors, namely Yin and Yang. The Yin-Yang duality is displayed not only in the stratum of cells, but also in a more deeper stratum - molecules (amino acids and nucleotides). In fact, in protein folding, the hydrophilic residues are exposed on the surface of globular protein but the hydrophobic residues burden in its interior. So, the hydrophilicity and hydrophobicity can be seen as a kind of Yin-Yang duality. On the other hand, the biosynthesis of protein is under the instruction of nucleotide sequence. It is unimaginable that if there is no existence of Yin-Yang duality in nucleic acids. On account of this, we propose the assumption of Yin-Yang duality of nucleotides. We emphasize the duality property of nucleotides and its relation to the characteristics and classification of codons and amino acids. In literatures, the similar model has been suggested by Swanson but from a different view of point [Swanson, 1984].

Four nucleotides are classified into purine and pyrimidine according to their chemical structure. They are classified into two kinds of Watson - Crick pairs according to the number of hydrogen bonds. So there exist the structural invariance between U and C or A and G and the symmetrical relation between U and A or C and G.. To express this symmetry we suppose that four nucleotides U,C,A,G are expressed by two lines (upper line and lower line) and each line takes two states , Yin denoted by — — and Yang denoted by —— . The two states of upper line are introduced to classify purine and pyrimidine, while that of lower line for sub-classification in purine or in pyrimidine. The W–C pair occurs between Yin and Yang. The representations are given by Figure 5a or equivalently by Figure 5b. In the following we will use the former representation of Figure 5a.

U and A are Yin-Yang symmetrical, C and G are Yin-Yang symmetrical, too. W- C bonds take place between them. The above representation of bases is called the assumption of duality or Yin-Yang of nucleotides. The representation is taken from 《*The book of Changes*》(《I Ching》) — a book of Chinese ancient philosophy. In this book Yin and Yang are introduced as two universal and fundamental properties — mutual contradictory and dependent with each other — of all things in nature and society. The four doublets of Yin and Yang are called four Yis which gives the detailed classification of Yin and Yang. U is called L-Yang (Lao-Yang or Large Yang); C, S-Yang (Shao-Yang or Small Yang); G, S-Yin (Shao-Yin or Small Yin) ; and A, L-Yin (Lao-Yin or Large Yin). The eight triplets of Yin and Yang are called eight Guas. Each Gua characterizes a phenomenon in nature. A pair of Guas (namely the hexameron of Ying and Yang ) describes a change (a changing state) of things. The above dual representation of nucleotides was introduced by us in 1992 [Luo, 1992]. But in a popular literature 《*Who wrote the book of life? – A history of the genetic code*》 [Kay, 2000] we read that "Around 1969 several individuals in Europe and the United States observed, from very different professional vantage points, that the ancient Chinese I

Ching and the newly completed genetic code shared remarkable similarity. The three-thousand-years old Book of Changes – a symbolic system for comprehending human experience – and the genetic Book of Life exhibited striking correspondence.” It seems that many people have noticed the similarity between the genetic code and the ancient Chinese I Ching. They take the nearly same view without prior consultation.

The dual representation of nucleotides reflects not only the intrinsic symmetry between four bases, but also the similarity order of them. U is placed in one end, A is placed in another end , C and G between them. The similarity between U and C (or A and G) is larger than that between U and G (or A and C), and the latter is, in turn, larger than U and A, since the upper line classifying purine and pyrimidine has a higher weight than the lower line. Two bases with large similarity will have high mutational rate between them. The observation on pseudo-genes mutation approves the supposition (see Table 4-3)[Li, 1997].

**Table 4-3 Relative mutational frequencies in pseudo genes**

| Mut to / Ori | A | T | C | G |
|---|---|---|---|---|
| A | – | $4.7 \pm 1.9$ | $5.2 \pm 0.8$ | $11.4 \pm 1.6$ |
| T | $4.5 \pm 1.0$ | – | $6.2 \pm 1.8$ | $4.6 \pm 1.8$ |
| C | $8.3 \pm 1.4$ | $22.0 \pm 1.8$ | – | $4.7 \pm 1.0$ |
| G | $16.0 \pm 1.1$ | $7.0 \pm 1.5$ | $5.5 \pm 0.8$ | – |

The representation is also comparable with the order of resistance of bases to ionizing radiation

$$A > G > C > U \geqslant T$$

(in presence of $O_2$, C and G may be transposed). The result is consistent with the theoretical calculation of resonance energy per π electron, $0.32\beta$, $0.27\beta$, $0.23\beta$, $0.19\beta$, $0.17\beta$ for A,G,C,U,T respectively [Pullman & Pullman, 1964]. Moreover, the representation is also consistent with the hydrophobicity order of nucleoside 5'- monophosphate, AMP > GMP > CMP > UMP, see Table 4-4 [Lacey & Mullins, 1983].

**Table 4-4 The hydrophilicity -*A* and hydrophobicity -*B* for nucleoside 5'- monophosphate**

| | AMP | GMP | CMP | UMP |
|---|---|---|---|---|
| *A* | 0.26 | 0.44 | 0.62 | 0.69 |
| *B* | 1.10 | 0.53 | 0.35 | 0.30 |

（*A* is taken from Weber & Lacey 1978；*B* - taken from Garel 1973. see Lacey & Mullins, 1983）

The definite order of four nucleotides – UCGA – is an important factor in understanding the broken symmetry. The base order has been changed from conventional UCAG to UCGA in above Ying-Yang representation. The point is also consistent with the structural regularity in nucleobases: the sp2 nitrogen atom number in nucleobase is 0 in U, 1 in C, 2 in G and 3 in A which was indicated by Yang [Yang, 2005].

The genetic code is triplets of nucleotides. Each codon should be represented by a diagram with 6 lines. We suppose that double lines corresponding to the first base are put in the center (the 3rd and 4th line of the six-line-diagram), double lines corresponding to the second base are put on its upper and lower sides (the 2nd and 5th line), and double lines corresponding to the third base are put on the exterior of the diagram (the 1st and 6th line). For example, tryptophan (Trp) is expressed by

— —
— —
——
——
——
——

Since the first two bases are more important in the determination of the property of amino acid we suppose that the central four lines are fundamental in the six-line-diagram. The genetic code table represented by central four lines of 64 codons is shown in Fig.6 .

Several symmetry operations can be defined as follows:

1) Star operation. By changing —— (– –) to – – (——) a codon X is transformed to its antonym (codon composed of complementary bases) X*;

2) R operation. By interchanging each upper line with the corresponding lower line a codon X is transformed to $X^R$. If $X^R$ =X it is called R-symmetrical; if $X^R$ =X* it is called R- antisymmetrical. The initiator and terminators UAA, UAG (first two bases) are R- symmetrical. CC, GG, CG, and GC belong to R-antisymmetrical.

3) T operation It is defined by interchange of lines (3,4) and (2,5) in six-line-diagram, that is, the interchange of first and second letter of a codon. The higher weight of the second position of a codon than the first position means the asymmetry under T operation.

Hydrophilicity and hydrophobicity are a kind of universal Yin-Yang duality of life expressed at the level of amino acids. In Fig.6 the genetic code has been represented through Yin line (— —) and Yang line (——). We find that the more Yin lines (— —) the di-nucleotide contains, the stronger the hydrophilicity of the encoded amino acid is; the more Yang lines (—) the di-nucleotide contains, the stronger the hydrophobicity of the encoded amino acid is. If the number of Yin lines is equal to that of Yang lines for a di-nucleotide, then the higher weight of upper lines as compared with lower ones and the higher weight of second position of a codon (2nd or 5th in six- lines) as compared with the first position (3rd or 4th in six-lines) should be considered. Following these rules we can divide the genetic code of Figure 6 into two regions. The amino acids inside the solid line (lower-right part of the figure) are hydrophilic and outside it (upper-left part) — hydrophobic. When the conventional order of bases, namely UCAG, has been changed to UCGA (the order of Yin-Yang), a very symmetric fashion of the hydrophilic - hydrophobic domain in the code table is obtained. The above hydrophilic - hydrophobic classification of amino acids is consistent with experimental data. (The case of di-nucleotide UC framed by dotted line should be considered carefully, that has been discussed in section 2.) So, the Yin-Yang duality provides a new explanation on the domain-like distribution of amino acids in the genetic code: The base A (Yin lines) in a codon contributes more to the amino acid hydrophilicity and the base U (Yang lines) contributes more to the amino acid hydrophobicity. The base G and C are in the middle of A and U [Luo, 1992; 2000; 2004]. In section 2 we have deduce hydrophilic – hydrophobic domain in the genetic code under the condition (1.10) and its complement (2.8). These inequalities on mutational parameters reflect the existence of some definite order about the base property among U,C,G and A. Now the base order and symmetry has been summarized by the formulation of Yin-Yang duality. Thus, the Yin-Yang duality can serve as a basic idea for understanding the hydrophilic – hydrophobic distribution of amino acids in the genetic code.

According to the proposed diagram representation, it is easily to find that the codon and its antonym behave differently in their Yin-Yang. So, if any amino acid, apart from Ser, is hydrophilic (hydrophobic), then the amino acid encoded by its antonym is hydrophobic (hydrophilic). In fact, the mutation rate between a codon and its antonym is very small since it is three-base mutation and it occurs between Watson - Crick pairs. By use of the similar method described in section 2 we can prove that a pair of codon and antonym are arranged in regions with different hydrophobicity.

Why the amino acid hydrophobicity can be deduced so successfully from the dual representation of nucleotides? The molecular mechanism is related to the tRNA structure and the origin of the genetic code. The selective interaction in the formation of tRNA molecule leads to the hydrophilic amino acid recognizing hydrophilic anti-codon, and the hydrophobic amino acid recognizing hydrophobic anti-codon. Considering that the base A in a codon contributes more to the amino acid hydrophilicity while the base U in a codon contributes more to the amino acid hydrophobicity, the base A (in an anticodon) should contribute more hydrophobicity to the dinucleoside monophosphate in anticodon and the base U (in an anticodon) should contribute more hydrophilicity to the dinucleoside monophosphate in anticodon. For example, dinucleoside monophosphate AA (Phe, Leu) and UU (Lys, Asn) have the lowest and highest hydrophilicity values 0.023 and 0.389 respectively. (see Table 4-4, where data on nucleoside 5'- monophosphate are listed, the similar data on dinucleoside monophosphate can be found in Lacey & Mullins, 1983).

The Yin-Yang duality affords a sound basis for understanding the hydrophilic – hydrophobic domain structure in the genetic code. It also provides an explanation on the robustness of the distribution under the variation of amino acids in the evolution.

Another important characteristic of amino acid is its volume (Table 4-5). They are roughly classified into two categories — the first ten are small amino acids while the last ten are large amino acids. The large amino acid is stiffer while the small one is more flexible. So, as the hydrophobicity, the volume of amino acid also plays an important role in protein folding, too. The volume classification of amino acids is shown in Figure 7, where codons encircled by solid lines code for small amino acids and those in the outer code for amino acids with large volume. Two kinds of amino acids classified by volumes are also located in separate domains in the code table [Luo, 1992].

**Table 4-5 Amino acid volume**

| | | | | | | | | | |
|---|---|---|---|---|---|---|---|---|---|
| Gly | Ala | Ser | Cys | Asp | Pro | Thr | Val | Asn | Glu |
| 3 | 14 | 21 | 30 | 30 | 31 | 32 | 36 | 36 | 41 |
| Ile | Leu | Gln | His | Met | Lys | Phe | Tyr | Arg | Trp |
| 46 | 46 | 47 | 50 | 52 | 58 | 62 | 69 | 70 | 83 |

（in unit of cubic angstrom，with a scale factor 2.01）

## 5 Conclusions

1. The synonym redundancy distribution in the genetic code is determined by the mutational parameters, the relative rates between transitional and transversional mutations and between 1-2 codon position and 3rd codon position mutations. The distribution is robust relative to the parameter choice. Under the constraints of $u$, $v$, $w_u$ and $w_v$ given by Eq (1.10) the pattern of codon degeneracy in the code can always be deduced.
2. The hydrophilic-hydrophobic domain in the genetic code is also robust under the mutational parameter choice and the variation of the distribution of amino acids in the code table. The robustness reflects the Ying-Yang duality existed among four nucleobases and 64 codons. The Ying-Yang duality emphasizes the definite order and the duality-symmetry among four nucleotides in codons.
3. MD theory gives an estimate on the accuracy of the genetic coding. The error of the genetic code comes from base mutation and translation. The two factors can be considered independently in the GMD formulation (Eq (2.1)). The mutational error can be minimized by the parameter choice based on the determination of synonym redundancy distribution and the translational error can be minimized through the appropriate arrangement of amino acids on the optimized hydrophilic-hydrophobic domain. In the proposed theory the optimal code is deduced through GMD minimization under the constraint of given amino acid number and given degeneracy degree for each amino acid. Apart from the estimation of the genetic coding accuracy, the GMD minimization reflects the selection process in the code evolution. The GMD is essentially a measure of non-fitness of the genetic code and the ideal code (expressed by [U] in Eq (2.1)) serves as the coordinate of Wright's adaptive landscape. The landscape changes, adaptive to the constraints on coordinate [U]. Then, the fittest code is selected out on the adaptive landscape. Therefore, in MD theory the genetic code origin is a problem of the evolution towards the optimal code (the fittest code) adiabatically on a given adaptive landscape if the landscape changes much slowly than the codon mutation and selection. The historical variation of the adaptive landscape (changed with the constraints on the degeneracy degree of each amino acid and the total number of encoded amino acids) is a central issue to be clarified for founding a comprehensive evolutionary theory.
4. However, from the preliminary calculation of GMD minimization under 20 amino acids with multiplicity distribution as in the standard code we find that under the initial fixation of some early amino acids on the code and under the doublet bundle of pairs of late amino acids with common precursor the standard code can be deduced logically. It shows the evolutionary accessibility of the prevalent standard code and may infer the big-bang-like formation of the standard code in a relatively short time after the Last Universal Common Ancestor of extant life (LUCA).
5. The mechanism for the evolvability of the prevalent standard code is mainly due to the alteration in tRNAs. The variation of optimal GMD can be calculated by using the local minima of MD which provides an approach to study the evolvability of the code. We find that, apart from the reassignment of a sense codon to a stop codon, the evolution of alternative genetic code has a general trend of GMD non-increasing and the finally observed reassignments are selective-advantageous or nearly neutral ones. Many ambiguities of the intermediate have been cleared up naturally through the selection role of MD minimization.

Acknowledgement The author is indebt to Dr Li Xiaoqin for her help in numerical calculation on the deduction of the minimal code table.

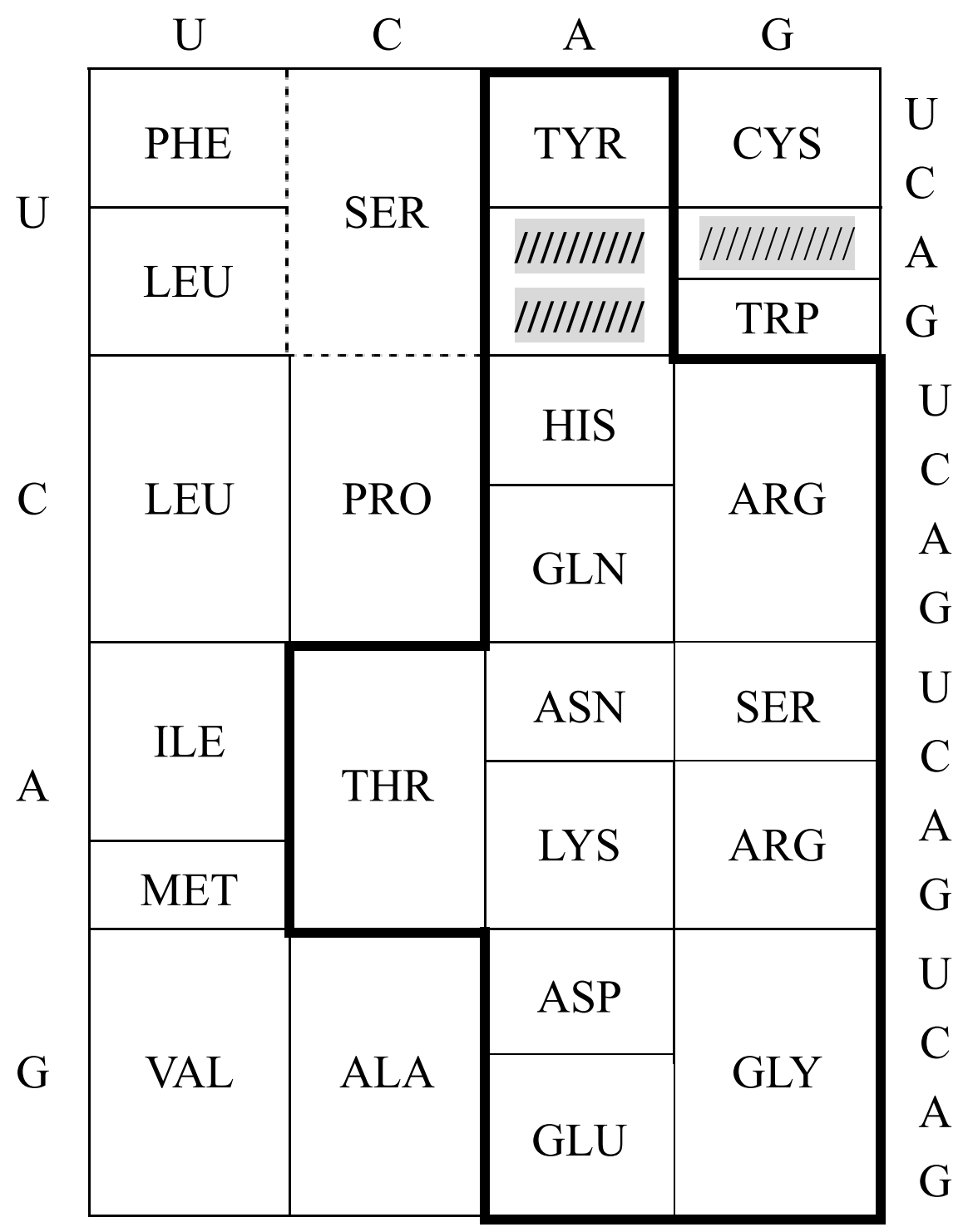

**Figure 1 The standard genetic code**

Amino acids inside the solid line are hydrophilic and outside it—hydrophobic. The domain-like distribution of amino acids in the code table is called hydrophobic-hydrophilic domain. For details see section 2.

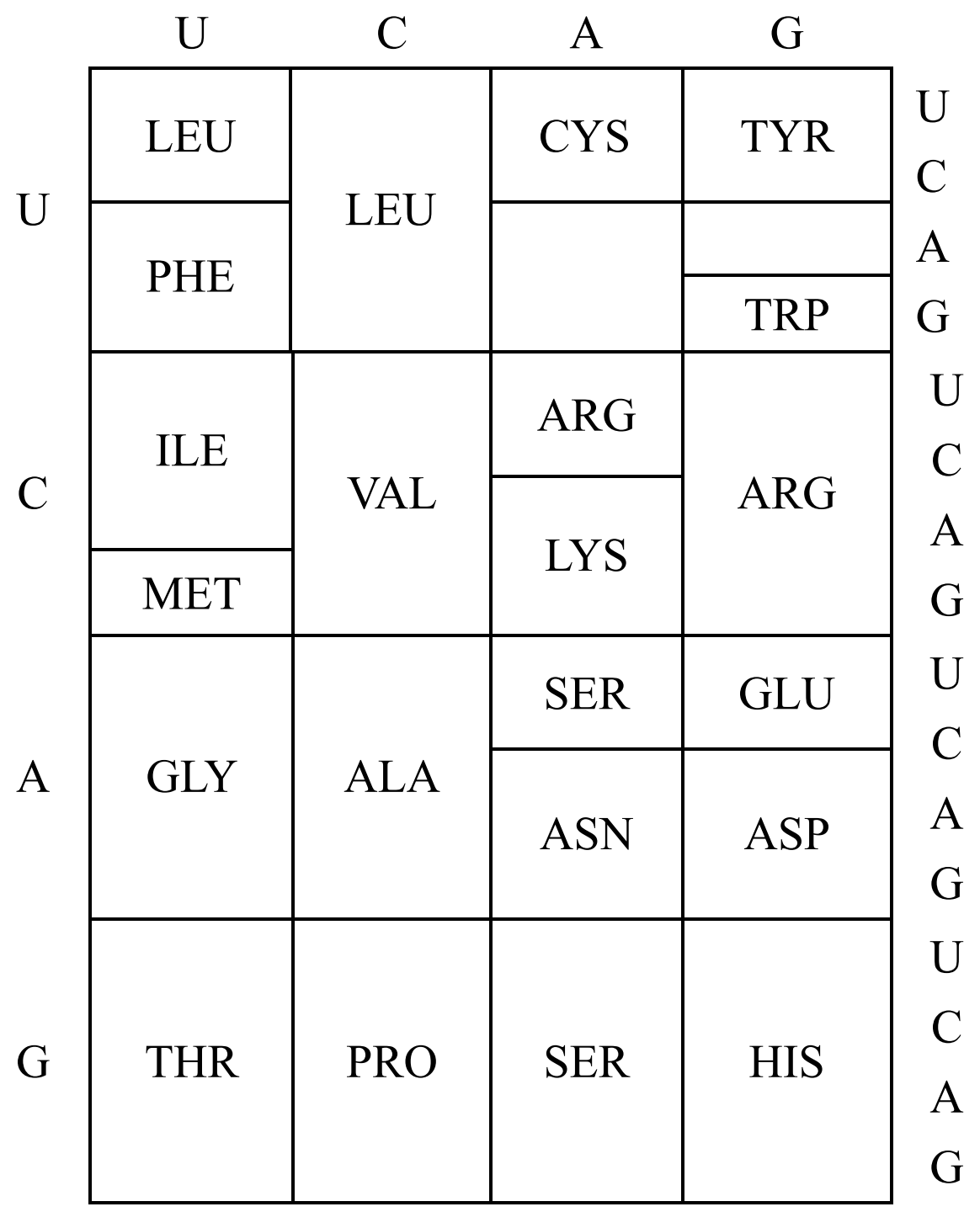

**Figure 2 The minimal genetic code deduced from minimization of GMD *Q'*(*U*)**

($Q_{min}$= 41722，see text, section 2)

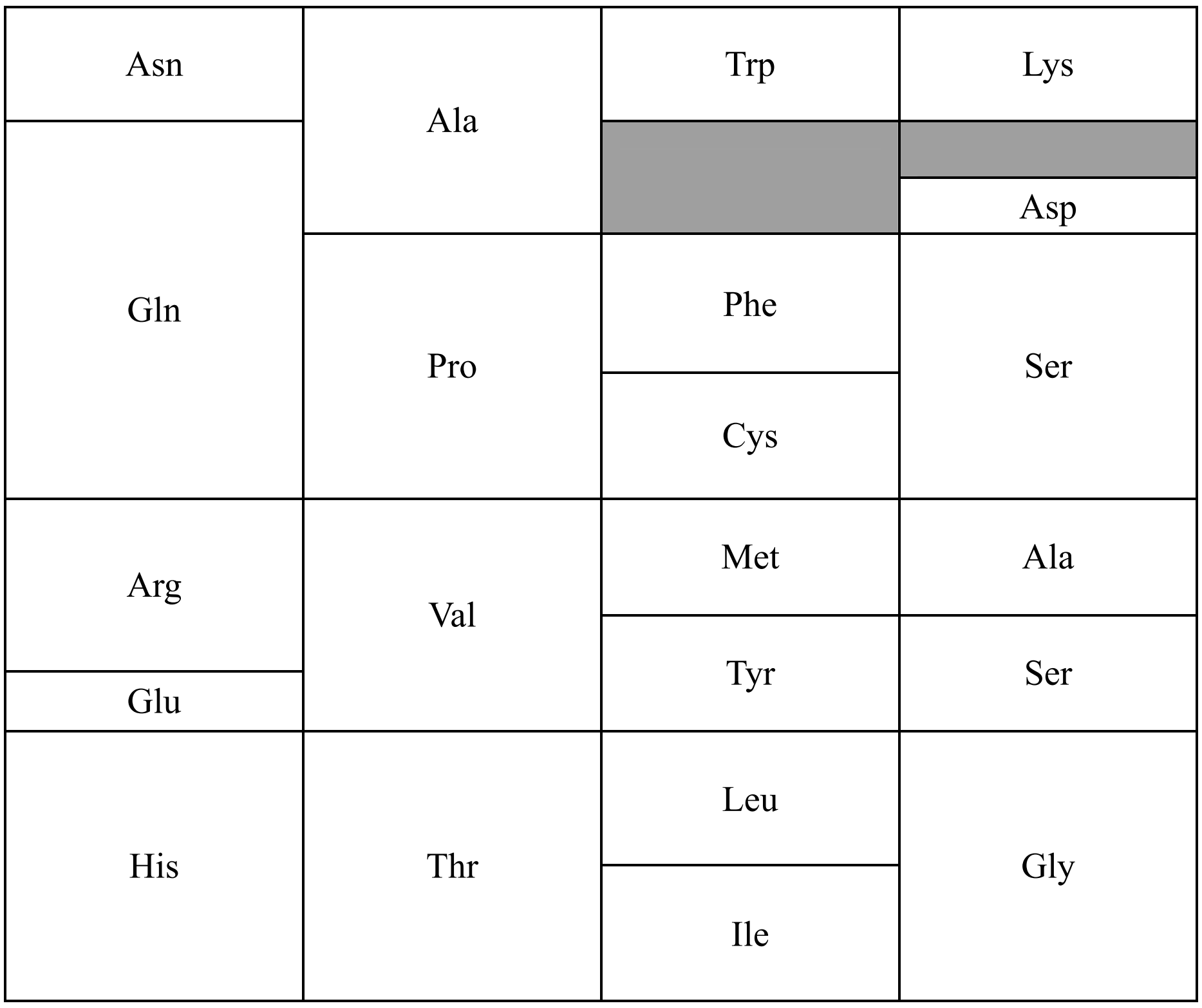

**Figure 3 The minimal code deduced by Di Giulio et al** (1994)

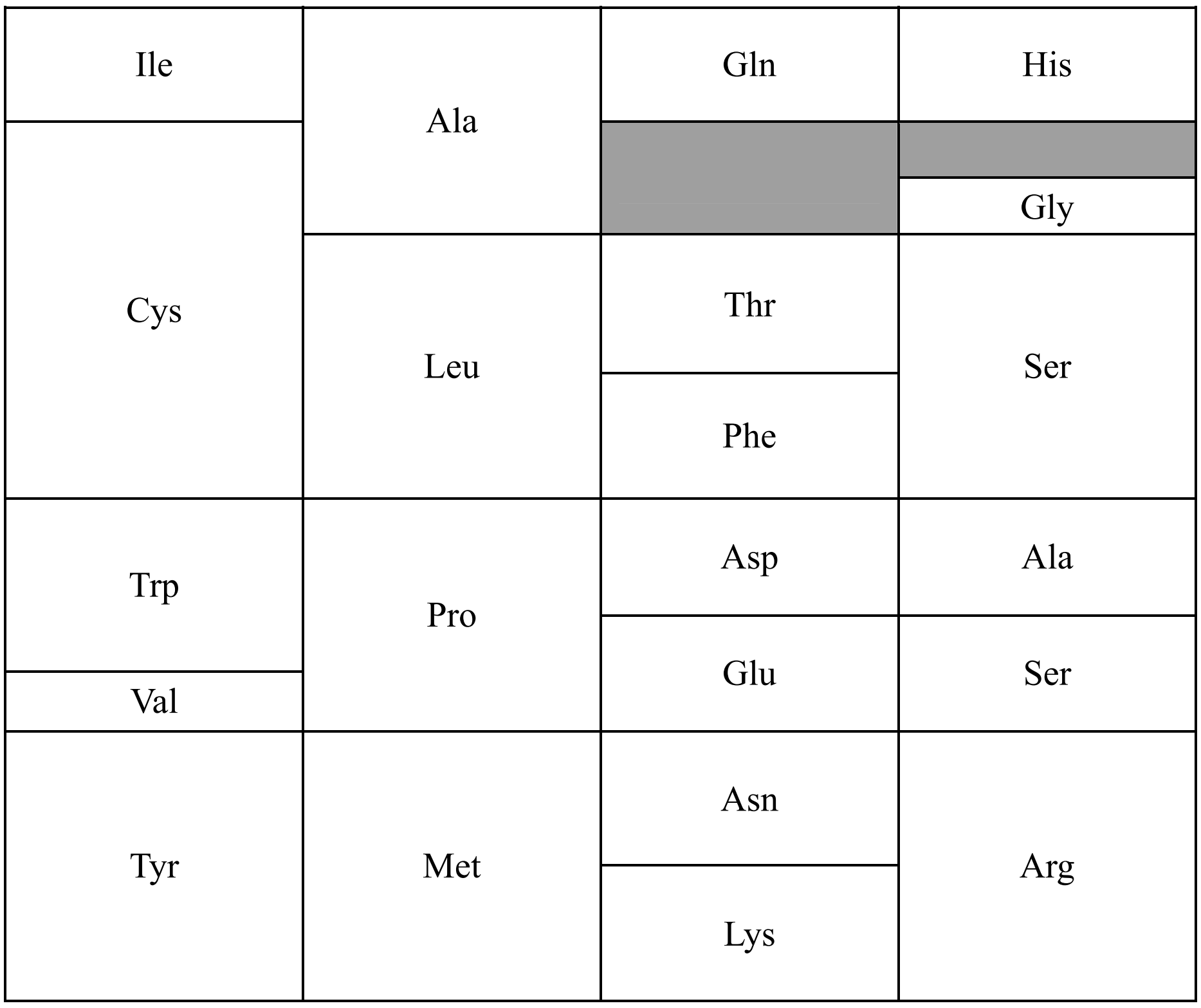

**Figure 4 The lower code deduced by Freeland and Hurst** (1998)

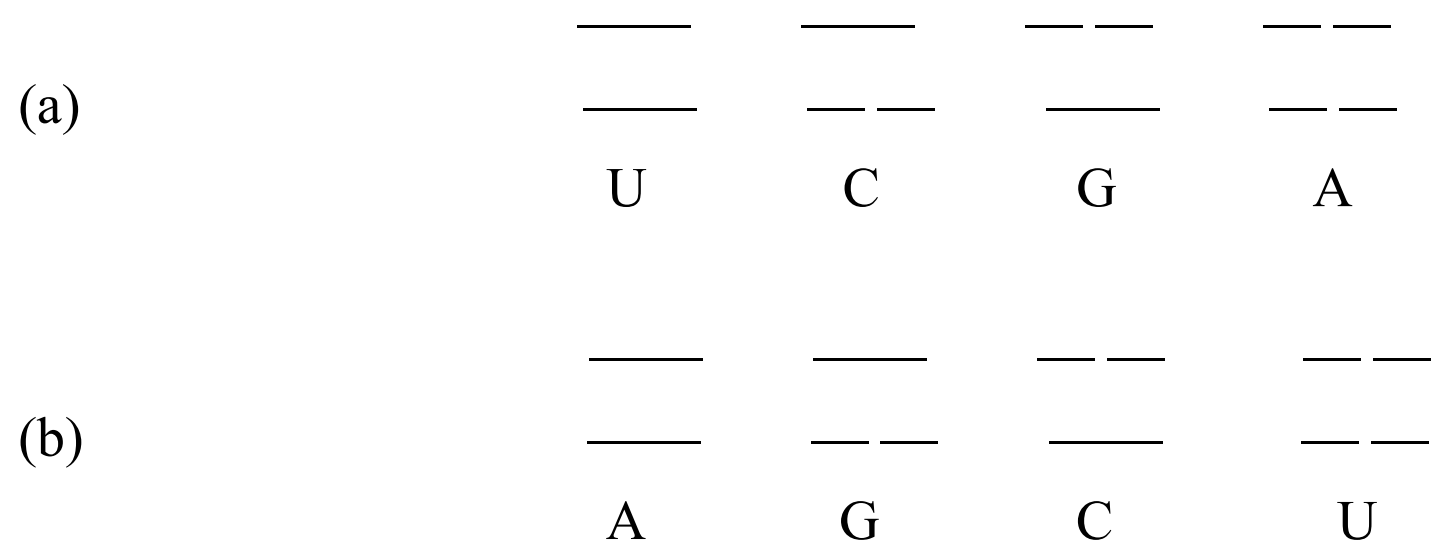

**Figure 5  The dual representation of nucleotides**

(see text, section 4)

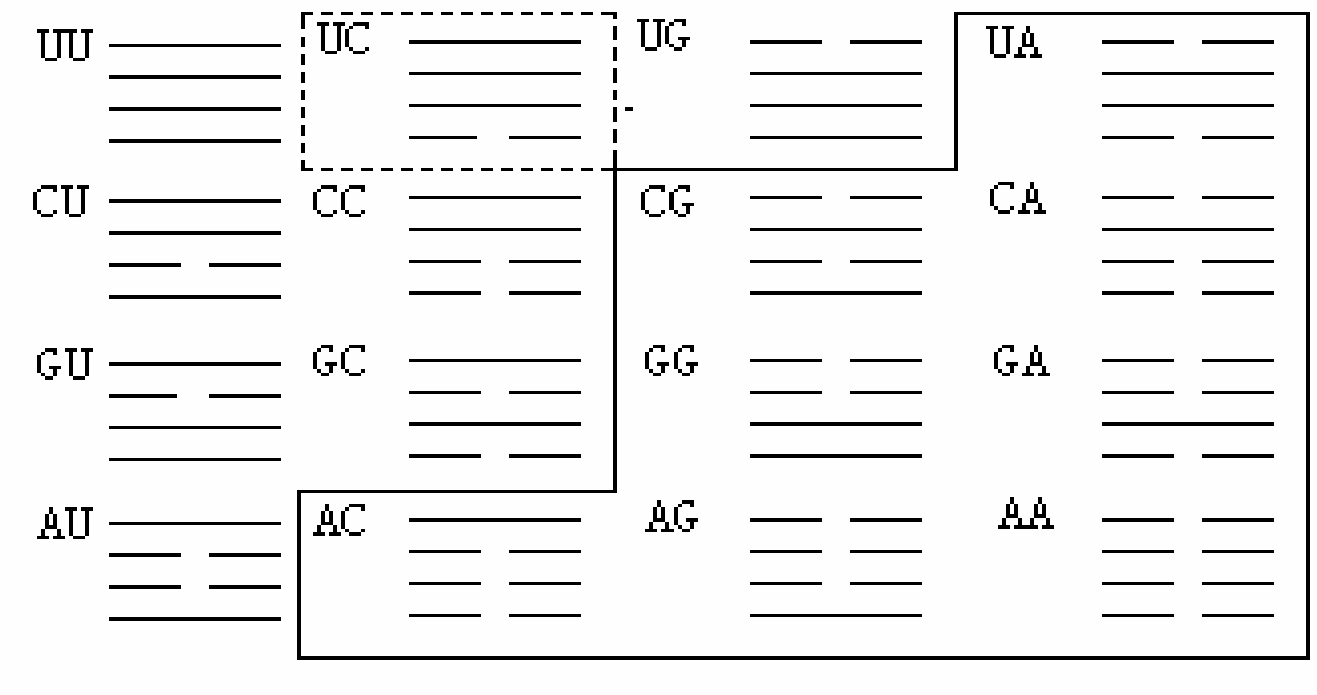

**Figure 6  The genetic code plotted with dual representation of nucleotides**

The base order has been changed to UCGA and a more symmetric hydrophobic-hydrophilic domain can be obtained in this order as shown in figure. Hydrophobic amino acids are located outside the solid line, and hydrophilic amino acids inside the solid line.

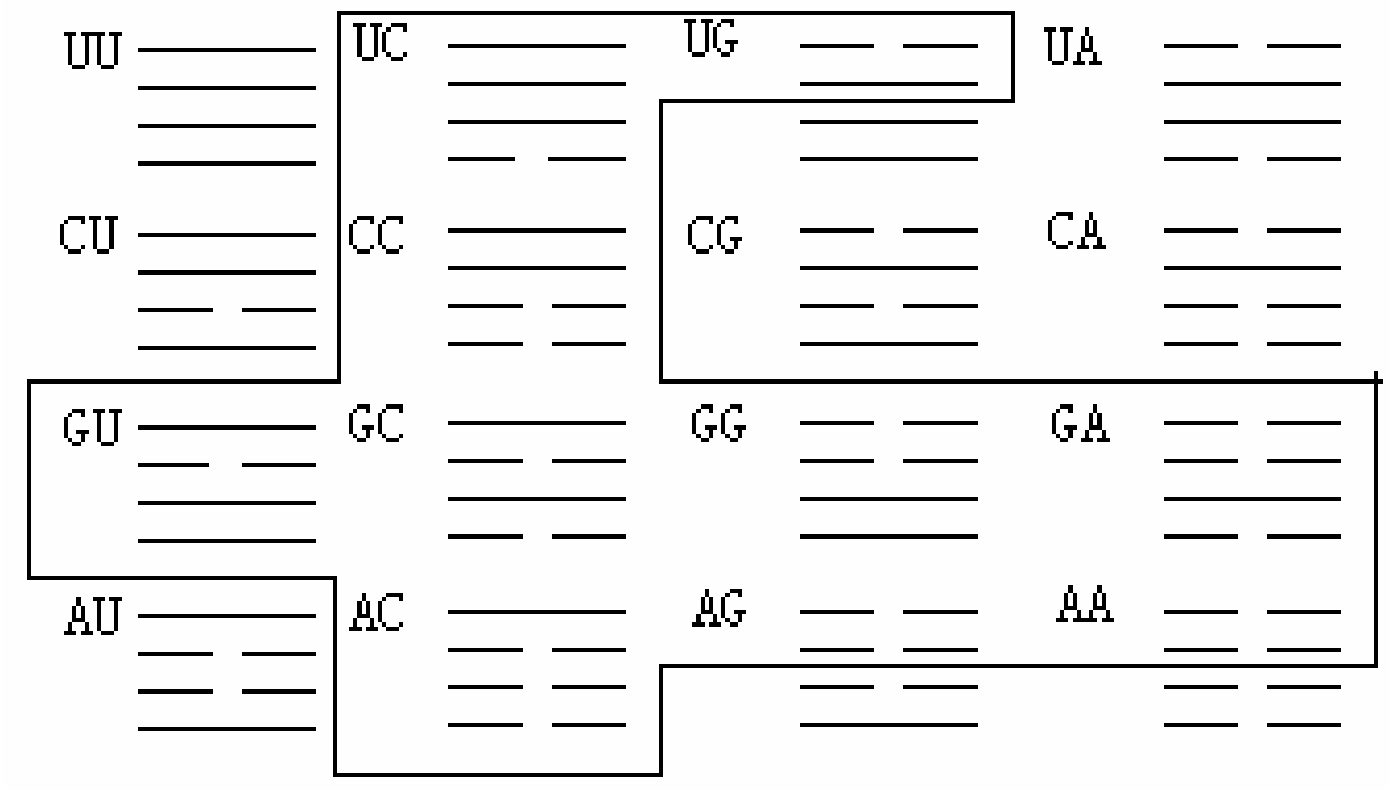

**Figure 7  The volume classification of amino acids**

The small amino acids are encircled by solid lines while the large amino acids located at their outer.

**Note Added in Publication**

The series of work on the origin and evolution of the genetic code was published from 1988 to 2004（see papers: Luo 1988; Luo 1989; Luo 1992; Luo 2000; Luo & Li, 2002a; Luo & Li, 2002b, and Luo 2004 listed in references）. It has been five years since the last paper in this series was published. However, we feel that till now the proposed theory, with its related calculation and research results in the series can still be regarded as one of the best theories on the genetic code evolution. Considering part of work was published in domestic journals and part of views was expressed in physical language unfamiliar to the circle of biologists, we combined these papers into one and rephrased it in a way which biologist may find easy to understand. All calculations have been checked but the results remain the same. Except for a few new papers are added into the references, most referred publications dated before 2004. We hope through the platform of open access, this theory and its research questions would attract wider attention.